# Forecasting the rates of future aftershocks of all generations is essential to develop better earthquake forecast models

**Shyam Nandan[1], Guy Ouillon[2], Didier Sornette[3], Stefan Wiemer[1]**

[1]ETH Zürich, Swiss Seismological Service, Sonneggstrasse 5, 8092 Zürich, Switzerland

[2]Lithophyse, 4 rue de l'Ancien Sénat, 06300 Nice, France

[3]ETH Zürich, Department of Management, Technology and Economics, Scheuchzerstrasse 7, 8092 Zürich, Switzerland

Corresponding author: Shyam Nandan (shyam4iiser@gmail.com)

**Key Points:**

1. Rigorous pseudo-prospective experiments reveal the forecasting prowess of the ETAS model with spatially variable parameters over the current state of the art smoothed seismicity models.

2. Smoothed seismicity models that account for the rates of future aftershocks of all generations can have superior forecasting performance relative to only background seismicity-based models.

3. Accounting for spatial variation of parameters of the ETAS model can lead to superior forecasting performance compared to ETAS models with spatially invariant parameters.

**Abstract:**

Currently, one of the best performing and most popular earthquake forecasting models rely on the working hypothesis that: "locations of past background earthquakes reveal the probable location of future seismicity". As an alternative, we present a class of smoothed seismicity models (SSMs) based on the principles of the Epidemic Type Aftershock Sequence (ETAS) model, which forecast the location, time and magnitude of all future earthquakes using the estimates of the background seismicity rate and the rates of future aftershocks of all generations. Using the Californian earthquake catalog, we formulate six controlled pseudo-prospective experiments with different combination of three target magnitude thresholds: 2.95, 3.95 or 4.95 and two forecasting time horizons: 1 or 5 year. In these experiments, we compare the performance of:(1) ETAS model with spatially homogenous parameters or GETAS (2) ETAS model with spatially variable parameters or SVETAS (3) three declustering based SSMs (4) a simple SSM based on undeclustered data and (5) a model based on strain rate data, in forecasting the location and magnitude of all (undeclustered) target earthquakes during many testing periods. In all conducted experiments, the SVETAS model comes out with consistent superiority compared to all the competing models. Consistently better performance of SVETAS model with respect to declustering based SSMs highlights the importance of forecasting the future aftershocks of all generations for developing better earthquake forecasting models. Among the two ETAS models themselves, accounting for the optimal spatial variation of the parameters leads to strong and statistically significant improvements in forecasting performance.

**1 Introduction**

With the recognition of our limited and fragmentary knowledge of the processes leading to seismogenesis, a prominent shift in the view of the seismological community has occurred, with a concentrated focus on earthquake forecasting instead of prediction [Jordan and Jones, 2010]. As a result, significant efforts have been invested in devising and validating models that yield probabilities of earthquake occurrence within given space-time-magnitude windows. The recent Regional

Earthquake Likelihood Models (RELM) experiment [Schorlemmer and Gerstenberger, 2007; Field, 2007], which is now being continued by the Collaboratory for the Study of Earthquake Predictability (CSEP) [Werner et al., 2010; Schorlemmer et al., 2010b; Zechar et al., 2010], serves as an excellent example of this concentrated effort. The success of RELM experiments and CSEP has been in bringing the earthquake forecasting research on the centre stage by providing a unique platform for systematically comparing predictive performances of various seismicity models. A significant fraction of the models considered during the first RELM experiments and by CSEP are smoothed seismicity models or SSMs, which rely on the working hypothesis that the "spatial location and magnitude distribution of the past earthquakes reveal the probable location and size of the future events". Different researchers have used this idea in different ways to construct coarse-grained forecasts of future earthquakes [Zechar et al., 2013; Schorlemmer et al., 2010a]. Some simply counted the total seismicity rate in predefined rectangular cells and scaled it appropriately to issue the forecasts for the required magnitude thresholds [Ebel et al. 2007; Wiemer and Schorlemmer, 2007]. Others employed relatively sophisticated kernel-based smoothing techniques of the location of past earthquakes to obtain a forecast of the future seismicity rate [Werner et al., 2011; Zechar and Jordan, 2010; Eberhard et al., 2012]. Some others even added an extra layer of sophistication by declustering the available training catalogues before smoothing and scaling [Helmstetter et al., 2007; Werner et al., 2011; Helmstetter and Werner, 2014; Kagan and Jackson, 2011]. The results of the RELM experiments [Zechar et al., 2013; Rhoades et al., 2014] have indicated that the smoothed seismicity approach based on the declustered earthquake catalogs (in particular, the model proposed by Helmstetter et al. [2007], which we will refer to as D-HKJ from now on) is the most informative not only when the competing models are other smoothed seismicity methods, but also when models use relevant strain and mapped fault information. However, one five-year experiment is not enough to reach stable conclusions about the best performing model. As a result, others have pursued the systematic testing and comparison of this smoothed seismicity approach. For instance, Stader et al. [2017] found that even though the model of Helmstetter et al. [2007] failed some consistency tests, particularly the N-test, the model still outperformed the physics-based models such as Uniform California Earthquake Rupture Forecast

(UCERF2) and National Seismic Hazard Mapping (NSHM) forecasts in several retrospective experiments with different time horizons.

Following the success of the D-HKJ model in the first RELM experiments, other researchers have proposed systematic improvements to this method, including incorporation of spatially variable estimates of b-values [Hiemer and Kamer, 2016], model hybridization [Rhoades et al., 2014; Rhoades et al., 2017; Steacy et al., 2013] and so on.

The success of SSMs that decluster especially compared to SSMs that do not, indicates that only the location of "background" earthquakes is important when forecasting the location of future large ($M \geq 5$) earthquakes at least on a time scale of 5 years. Indeed, this requires us to reformulate the aforementioned working hypothesis to: "locations of the past background earthquakes reveal the probable location of future seismicity".

In this paper, we explore an alternative class of smoothed seismicity models that are based on the principles of the Epidemic Type Aftershock Sequence (ETAS) model, which have been widely used in the literature to model short-term (on order of days to weeks) fluctuations in the seismicity rate following a large earthquake [Werner et al., 2011; Console et al., 2007; Iwata, 2010; Zhuang et al., 2002; Helmstetter and Sornette, 2002a, 2002b, 2003a and 2003b; Helmstetter et al., 2003; Daley and Vere-Jones, 2002; Ogata, 2011]. We aim to investigate if the same model can be extended to make forecasts for longer forecasting horizons (5 years, for instance, as in the RELM experiments). The ETAS models differ from the declustering based SSMs in the key idea that they not only utilise the estimates of the background seismicity rate based on the past catalogue but also parametric equations to forecast the rates of future aftershocks. We consider two versions of the ETAS model: one using a global set of parameters, and another one with spatially variable parameters. We have already proposed an objective framework to obtain spatially variable parameters of the ETAS model in Nandan et al. [2017], where we justified the parametric complexity of this model using the Bayesian Information Criterion. However, it remains to be checked if this "optimally" complex ETAS model can indeed outperform its more straightforward counterpart in a controlled pseudo-prospective experiment. For reference, we also compare the performance of these ETAS based models to three declustering based SSMs [Werner et al., 2011; Helmstetter et al., 2007; Zaliapin et al., 2008], a simple SSM based on undeclustered past earthquake catalog [Zechar and Jordan, 2010], and a model based on strain rate data [Bird and Liu, 2007]. In order to ascertain if the derived conclusions in this study are robust with respect to the starting year of the testing period (1999, 2000,…,2016), its duration (1 or 5 year) and minimum magnitude threshold of target earthquakes (2.95, 3.95 or 4.95), we perform numerous pseudo-prospective experiments with different combinations of these "hyper parameters". In these experiments, the competing models use the earthquake data during the training period to forecast the location and magnitude of all earthquakes, not just some selected earthquakes based on predefined declustering criteria, during the testing period defined using the above mentioned hyper parameters.

The paper is organized as follows. In section 2, we describe the data used in this study along with other essential and necessary considerations before calibration of the SSMs. Then, in section 3, we describe in detail the formulations of the six competing models. Further, we present and discuss the results of the pseudo-prospective experiments conducted in section 4. Finally, section 5 outlines our conclusions. As this paper relies on extensive use of acronyms, we have given the provided the definition of all the important acronyms in Table 1.

**2 Data**

The pseudo-prospective experiments undertaken in this study focus on testing the performance of the smoothed seismicity models in and around the state of California. We use the earthquake catalogue from the Advanced National Seismic System (ANSS) database that includes earthquakes in this region (see Figure 1). More specifically, we use the collection polygon defined for the Regional Earthquake Likelihood Model (RELM) experiment [Schorlemmer and Gerstenberger, 2007] as the spatial boundary of the catalogue.

Completeness of the earthquake catalogue is an essential factor for the construction of the smoothed seismicity models. Its influence is two folds. First, it affects the calibration of the models, thereby influencing the rate forecasts that are obtained. For instance, in the case of ETAS based models, incorrect estimates of the magnitude of completeness ($M_c$) can lead to biased estimates of parameters and consequently to biased rate forecasts [Seif et al., 2016]. Second, if one aims to forecast the rate of seismicity at any specific target magnitude threshold (say $M_t$), using the earthquakes larger than $M_c$ in the training catalog, the smoothed seismicity forecasts need to be scaled by a factor of $10^{-b(M_t - M_c)}$, where $b$ is the exponent of the Gutenberg-Richter (GR) law. The scaling essentially assumes that the GR law is valid above the chosen magnitude of completeness. For an incorrect choice of $M_c$ for the training catalog, the scaling may lead to inappropriate forecasts. For the area under study, Nandan et al. [2017] found that the ANSS catalog is complete for $M \geq 2.95$ during the years 1981 to 2016. This choice of $M_c = 2.95$ is justified in Nandan et al. [2017] by estimating independent temporal and spatial variation of magnitudes of completeness in the study region. However, it is important to note that even though the choice of $M_c = 2.95$ is rather conservative for most of the study region and time periods, some important regions such as offshore Mendocino and some important time periods such as several days following the largest earthquakes in the catalog may violate this assumption. One way to remedy to such a problem could be to use $M_c$ values larger than 2.95. However, it comes at the cost of much less resolved smoothed seismicity models due to a lack of coverage resulting from the dwindling number of earthquakes with an increasing magnitude of completeness. Thus, we fix the magnitude of completeness to 2.95 for all forecasting experiments conducted in this study.

For all the models appraised here, we also considered an auxiliary catalogue ($M \geq 2.95$), within the collection polygon and time period 1971 to 1981 (see Figure 1). The role of earthquakes in this auxiliary catalog is to act only as parents of the earthquakes in the primary catalog. The consideration of such an auxiliary catalog is especially relevant for the calibration of ETAS based models (see Wang et al., [2010] for instance), to avoid the situation that earthquakes in the early part of the primary catalog might systematically appear as orphans (i.e. background events) due to the artificial lack of any ancestor.

## 3 Method

### 3.1 Pseudo-prospective forecasting experiments:

We perform 6 sets of pseudo-prospective tests. In each of these experiments, we define a training period and correspondingly a training catalogue which is used to construct forecasts for the testing period. The testing period is always a period immediately following the training period. Earthquakes in the testing period are used to evaluate the performance of the forecasts constructed using different models. In all the forecasting experiments conducted in this study, the earthquakes within the predefined Collection polygon are used for model construction. The collection polygon encloses a smaller predefined polygon called a testing polygon, which is divided into $0.1°\ X\ 0.1°$ grid cells. The forecasts of all models during any testing period are issued on these testing grid cells in terms of expected number of earthquakes during the predefined testing period in 0.1-unit magnitude bins. To be strictly consistent with the RELM experiments, we use the same grid centers (7682 in total) as used by Zechar et al. [2013]. In the following, we give a detailed set-up of the 6 experiments conducted in this study.

1. **Experiment 1**: The five-year-long experiments start with the testing period [1999, 2003]. This testing time interval is then systematically shifted by one year to [2000, 2004] and so on until the final testing period [2012, 2016] is reached. With each one-year shift of the testing period, we increase the duration of the training catalogue by one year. In this experiment, the different models forecast the rate of earthquakes with $M_t \geq 4.95$ in 0.1-unit magnitude bins.

2. **Experiment 2**: Same as Experiment 1 except that the models forecast the rate of earthquakes with $M_t \geq 3.95$ in 0.1-unit magnitude bins.

3. **Experiment 3**: Same as Experiments 1 and 2 except that the different models forecast the rate of earthquakes with $M_t \geq 2.95$ in 0.1-unit magnitude bins.

4. **Experiment 4**: These one-year long experiments start with the testing period [1999, 2000). Note that the different closing bracket indicates that the year 2000 is not included in the testing period. This testing time interval is then systematically shifted by one year to [2000, 2001) and so on until the final testing period [2016, 2017) is reached. With each one-year shift of the testing period, we increase the duration of the training catalogue by one year. In this experiment, the different models forecast the rate of earthquakes with $M_t \geq 4.95$ in 0.1-unit magnitude bins.

5. **Experiment 5**: Same as Experiment 4 except that the different models forecast the rate of earthquakes with $M_t \geq 3.95$ in 0.1-unit magnitude bins.

6. **Experiment 6**: Same as Experiment 4 and 5 except that the different models forecast the rate of earthquakes with $M \geq 2.95$ in 0.1-unit magnitude bins.

The performance of the models is evaluated based on the true location and magnitude of the earthquakes that occurred during the testing period, which are shown in Figure 1.

### 3.2 Competing models

We use four classes of competing smoothed seismicity models: (i) ETAS based models; (ii) declustering based models; (iii) a simple smoothed seismicity model and (iv) a model based on strain data. For all these models, we assume that the magnitudes of the forecasted earthquakes follow the GR law with a uniform b-value (b=1).

#### 3.2.1 ETAS based models:

#### 3.2.1.1 ETAS model with a global set of parameters (GETAS model)

In the ETAS model, the seismicity rate, $\lambda(t, x, y|H_t)$, at any time $t$ and location $(x, y)$ is conditional upon the history of seismicity, $H_t$, up to $t$. $\lambda(t, x, y|H_t)$ is explicitly formulated as:

$$\lambda(t, x, y|H_t) = \mu + \sum_{i: t_i < t} g(t - t_i, x - x_i, y - y_i, m_i) \quad (1)$$

In the above formulation, $\lambda(t, x, y|H_t)$ depends on two components: the background intensity function $\mu$, and the triggering function, $g(t - t_i, x - x_i, y - y_i, m_i)$, which is explicitly formulated as:

$$g(t - t_i, x - x_i, y - y_i, m_i) = \frac{K \exp[a(m_i - M_c)]\{t - t_i + c\}^{-1-\omega}}{\{(x - x_i)^2 + (y - y_i)^2 + d \exp[\gamma(m_i - M_0)]\}^{1+\rho}} \quad (2)$$

$\{\mu, K, a, c, \omega, d, \gamma, \rho\}$ is the set of parameters that characterise the ETAS model. We obtain the global estimate of these parameters using the Expectation Maximization (EM) formulation proposed by Veen and Schoenberg [2008]. On calibrating the GETAS model on the primary earthquake catalogue

(shown in Figure 1) covering the entire period (1981-2017), we find: $\mu = 2.31 \times 10^{-7}$ earthquakes km$^{-2}$ day$^{-1}$, $K = 3.95 \times 10^{-4}$, $a = 1.60$, $c = 1.74 \times 10^{-2}$ day, $\omega = 0.17$, $d = 0.24$ km$^2$, $\gamma = 1.11$, $\rho = 0.53$.

However, to ensure the sanctity of the pseudo-prospective experiments, the process of calibration of the ETAS model is repeated for all the training catalogues.

### 3.2.1.2 ETAS model with spatially variable parameters (SVETAS model)

The parameters of the ETAS model defined in the previous section may or may not feature spatial variation, but the former seems more likely considering that these parameters are possibly the manifestations of the local physical properties of the crust, which exhibit spatial variability. Nevertheless, these parameters have been generally assumed to be spatially invariant in most of the literature on the ETAS model, primarily for reasons of computational simplicity.

Recently, we introduced an objective method to invert spatially variable parameters of the ETAS model [Nandan et al., 2017]. We found that the three parameters $K$, $a$ and $\mu$ show substantial spatial variation, which could be explained by the spatial variation of the surface heat flow in the study region. In Nandan et al. [2017], however, we considered the parameters of the temporal (Omori) and spatial kernels to be spatially invariant for reasons of computational and mathematical simplicity. In the present work, we have extended our previous method to invert the spatial variation of the Omori parameters as well, while still assuming the parameters of the spatial kernel to be spatially invariant for reasons of computational simplicity. In the following, we give a summary of the inversion algorithm that we use to obtain the spatial variation of the ETAS parameters and adopt the same representation as in Nandan et al. [2017].

1. We first assume that the number of spatial Voronoi partitions (i.e. cells) that are necessary to capture the spatial variation of the parameters $\{\mu, K, a, c, \omega\}$ is q. In each of these q partitions, $S = \{S_1, S_2, S_3, \ldots, S_q\}$, the parameters $\{\mu, K, a, c, \omega\}$ are assumed to be piecewise constant. So, each partition $S_{i:1 \text{ to } q}$ has its own set of constant parameters $\{\mu, K, a, c, \omega\}_{i:1 \text{ to } q}$. The parameters $\{d, \gamma, \rho\}$ are the same in all the partitions. To divide the study region into q partitions, we draw q points without repetition from the set of known locations of earthquakes in the catalogue to be used as centres of the Voronoi partitions, while ensuring that within each of the resulting spatial partitions there are at least 100 earthquakes. This limit of 100 earthquakes is imposed to ensure that estimated parameters of the ETAS model are stable. Note that drawing the Voronoi centres from locations of earthquakes in the catalogue, rather than randomly choosing them within the study region, allows for a higher resolution in areas that are densely populated by seismicity. The modified sampling technique is thus more natural and data-driven.

2. We then use the extended EM algorithm proposed by us [Nandan et al., 2017] to invert for the parameters $\{\mu, K, a, c, \omega\}$ is each of the spatial partitions along with the global estimates of the parameters $\{d, \rho, \gamma\}$.

3. We then compute the penalised log-likelihood score for the model, $BIC = -2LL + N_{pars} \log N$, where $N_{pars} = 7q + 3$, as each spatial partition is characterised by one parameter from the background seismicity rate, two productivity parameters, two Omori parameters and two Voronoi centre parameters, along with three global parameters corresponding to the spatial kernel. Note that LL and N respectively correspond to the expected maximum log-likelihood score obtained upon the calibration of the ETAS model using the EM algorithm and to the number of earthquakes used for the calibration of the ETAS model.

4. We then repeat the above mentioned steps 1000 times with different realisations of q Voronoi centres selected randomly from the list of the earthquakes and store the estimate parameters and BIC.

5. We then repeat steps 1 to 4 with a value of q increasing from 1 to 150.

Having calibrated 150,000 models with varying degrees of complexity, we then rank these models according to their BIC and select the top 1% solution for computing an ensemble model, while giving the $i^{th}$ model out of the M selected models the weight of $\frac{\exp\left[-\frac{BIC_i}{N}\right]}{\sum_{i=1}^{M} \exp\left[-\frac{BIC_i}{N}\right]}$.

On calibration of the SVETAS model on the primary earthquake catalogue (shown in Figure 1) covering the entire period (1981-2017), we find the ensemble estimates of the global parameters: $d = 0.20 \text{ km}^2, \gamma = 1.21, \rho = 0.58$. In Figure S1, we show the spatial variation of the five other ETAS parameters. We replace the spatially variable estimate of the parameter K by the branching ratio (n), which is a more informative parameter. It quantifies the average number of direct aftershocks triggered by any earthquake averaged over all sizes.

Multitudes of factors can contribute to the spatial variation of the parameters. Some plausible ones include (i) stationary physical processes that genuinely vary in space due to their dependence on local rock composition, heat flux, water content and fault structures; (ii) spatial variation of the location uncertainty; (iii) variable anthropogenic forcing, and so on. While it would be essential to distinguish these effects from each other, this paper only intends to test the meaningfulness of these obtained estimated spatial variation regarding improved forecasting potential (if any) over the ETAS model with spatially homogeneous parameters.

Note that the above procedure of calibration of the SVETAS model is repeated for all the training catalogues to maintain the sanctity of the pseudo-prospective forecasting experiment.

### 3.2.1.3 Smoothed seismicity forecast using the GETAS and the SVETAS model

Having calibrated the two ETAS models on the training catalogue, a smoothed seismicity forecast for the corresponding testing period can be constructed by adding the contribution of the following three components:

A. **Spontaneous earthquakes occurring at rate $\mu$**

The contribution of exogenous processes is encoded in the parameter $\mu$. However, instead of directly using it for forecasting the future background seismicity rate, we follow a strategy akin to that of Helmstetter et al. [2007]. The calibration of the SVETAS model on the training catalogue using the EM algorithm yields a vital quantity: the independence probability $IP_i$ corresponding to each earthquake present in the training catalogue. $IP_i$ quantifies the probability that the $i^{th}$ earthquake in the training catalogue has not been triggered by a previous one. To construct the forecast of the background seismicity for the testing period, we first obtain a smoothed approximation of the background seismicity rate within the training period, by smoothing the location of the earthquakes present in the training catalogue. This can be achieved by using isotropic Gaussian kernels that are weighted according to the independence probability of the earthquakes. In a given spatial bin (pre-defined by CSEP) with index l and spatial extent $S_l$, the smoothed estimate of the number of background earthquakes during the training period is given by:

$$\widehat{N}_{bkg} = \sum_i \int_{S_l} k_i(x,y)\, dx\, dy$$
$$k_i(x,y) = \frac{IP_i}{2\pi\sigma_i^2} \exp\left[-\frac{1}{2}\left\{\frac{(x-x_i)^2 + (y-y_i)^2}{\sigma_i^2}\right\}\right] \quad (3)$$

$k_i(x,y)$ is a 2D isotropic Gaussian kernel with bandwidth $\sigma_i$, weighted by the independence probability, $IP_i$, of the $i^{th}$ earthquake in the training catalogue. In Equation (3), $x_i$ and $y_i$ represent the location of the $i^{th}$ earthquake in the catalogue; $\sigma_i$ is taken as the distance of the $j^{th}$ nearest neighbour background earthquake from the $i^{th}$ earthquake. For all the investigations conducted in this study, we have fixed j equal to 1.

Note that the analytical value of the integral $\int_{S_l} k_i(x,y)\, dx\, dy$ for the $i^{th}$ earthquake in the rectangular grid cell, with x and y limits respectively being $[x_1, x_2]$ and $[y_1, y_2]$, is given by the following expression [Zechar et al., 2010]:

$$\int_{x_1, y_1}^{x_2, y_2} k_i(x,y)\, dx\, dy = \frac{1}{4}\left[\text{erf}\left\{\frac{x_i - x_2}{\sigma_i\sqrt{2}}\right\} - \text{erf}\left\{\frac{x_i - x_1}{\sigma_i\sqrt{2}}\right\}\right]\left[\text{erf}\left\{\frac{y_i - y_2}{\sigma_i\sqrt{2}}\right\} - \text{erf}\left\{\frac{y_i - y_1}{\sigma_i\sqrt{2}}\right\}\right]$$

where erf(.) is the error function. Having obtained the smoothed number of background earthquakes in each of the spatial bins during the training period, we construct the forecast of the number of background earthquakes during the testing period by first normalizing the spatial density of the background earthquakes such that it integrates to 1 over the predefined testing polygon and then by scaling the spatial PDF by a factor $\frac{T_{test}}{T_{train}}\sum_i IP_i$. This upscaling ensures that the forecasted number of the background earthquakes is proportional to the time duration of the testing period. Note that the index i runs over only those earthquakes in the training catalogue that are enclosed within the testing polygon, while $T_{test}$ and $T_{train}$, respectively, are the duration of the testing and the training periods.

**B. Aftershocks of type I**

These aftershocks include not only the direct descendants of the earthquakes in the training catalog, but also the cascade of earthquakes that would be activated by those direct aftershocks. While computing the expected number of direct aftershocks in the testing period due to the earthquakes in the training catalogue is rather trivial and can be done analytically, one needs to perform simulations to compute the average number of higher generation aftershocks in the finite sized testing period. This is done using the following simulation scheme:

1. Simulate the first-generation aftershocks for all the earthquakes in the training period.

    a. An earthquake with magnitude $m_i$ can trigger on average $K_i\, e^{[a_i(m_i - M_c)]}$ aftershocks with magnitude larger than $M_c$, where $M_c$ is the magnitude of completeness. $K_i$ and $a_i$ represent the ensemble estimate of the productivity parameters at the location of the $i^{th}$ earthquake.

    b. Times of those aftershocks can be simulated according to the following equation:

    $$t_{ji} = c_i \left\{ (1 - U_j^1)^{-\frac{1}{\omega_i}} - 1 \right\} + t_i \qquad (4)$$

    In the above equation, $t_{ji}$ is the time of the $j^{th}$ aftershock that has been triggered by the $i^{th}$ earthquake in the training catalogue, which is counted from the time of the $i^{th}$ earthquake; $U_j^1$ is a random number drawn uniformly in the interval [0,1]; $c_i$ and $\omega_i$ are the ensemble estimate of the parameters of the Omori kernel at the location of $i^{th}$ earthquake that occurred at time $t_i$.

    c. Similar to the time of the aftershocks, we simulate the Cartesian coordinates of the direct aftershocks $\{x_{ji}, y_{ji}\}$ of the mainshock $\{m_i, x_i, y_i\}$ using the following equations:

    $$r_{ji} = \left[ d\, \exp\{\gamma(m_i - M_c)\} \left\{ (1 - U_j^2)^{-\frac{1}{\rho}} - 1 \right\} \right]^{\frac{1}{2}}, \quad \theta_j = 2\pi U_j^3 \qquad (5)$$
    $$x_{ji} = r_{ji} \cos\theta_j, \quad y_{ji} = r_{ji} \sin\theta_j$$

In the above equation, $r_{ji}$ and $\theta_j$ are respectively the simulated distance and the azimuth of the $j^{th}$ aftershock from its mainshock with index i; $U_j^2$ and $U_j^3$ are random numbers drawn uniformly in the interval [0,1]; d, γ and ρ are the ensemble estimates of the spatially invariant parameters of the spatial kernel.

d. We then simulate the magnitude of the direct aftershocks using the following equation:

$$m_j = \frac{-\log(1 - U_j^4)}{\beta} + M_c \tag{6}$$

In the above equation, $U_j^4$ is a number drawn uniformly at random from the interval [0,1] and $\beta = \log(10)$ is the assumed global exponent of the GR law.

2. Having simulated the direct aftershocks of all the earthquakes of the training catalogue, we then reject all those direct aftershocks that do not fall in the testing period.

3. All the direct aftershocks selected in step 2 are then treated as new mainshocks that are then allowed to trigger their own direct aftershocks whose time, location and magnitude are simulated using the steps a-d defined under step 1. Again, the direct aftershock selection criteria outlined in step 2 is applied.

4. Steps 2 and 3 are repeated until no more aftershocks are simulated.

5. To obtain the smooth coarse-grained map of the number of aftershocks (direct/indirect) triggered by the earthquakes in the training catalog, we perform numerous (300,000) simulations; we count the number of aftershocks that occurred in each of the predefined CSEP bins in each simulation and find the number of aftershocks in each of the CSEP bins averaged over all simulations.

C. **Aftershocks of type II**

One also needs to take account of the expected contribution of the direct and higher generation aftershocks triggered by the background earthquakes expected to occur in the testing period. The first step in the simulation of these aftershocks is to generate the background/seed events from which these aftershocks would emanate. This can be done by first drawing $U_i$, a number uniformly drawn at random from the interval $\left[0, \frac{T_{train}}{T_{test}}\right]$ and comparing it to the independence probability $IP_i$ of the $i^{th}$ earthquake in the training catalog. If $U_i \leq IP_i$, then the location of the $i^{th}$ earthquake is selected as the location of a new background event. Note that generating the random number between $\left[0, \frac{T_{train}}{T_{test}}\right]$ ensures that the expected number of background earthquakes during the testing period is $\frac{T_{test}}{T_{train}} \sum_i IP_i$. This process is repeated sequentially for all the earthquakes in the training catalogue. For the selected seed events, we discard their actual time and magnitude information and only retain their location information. These events are assigned times that are distributed uniformly randomly between $[T_{train}, T_{train} + T_{test}]$ and the magnitudes of these events are simulated as for the Type I aftershocks. We then simulate the cascade of aftershocks corresponding to all the simulated seed events as described for Type I aftershocks. Finally, we obtain a smooth coarse-grained map of the number of aftershocks (direct/indirect) triggered by the background earthquakes in the testing period by performing numerous (300,000) simulations and finding the average the number of aftershocks in each of the CSEP bins in all simulations.

### 3.2.2 Background seismicity-based models

While the earthquakes that occur in nature do not come with the label of "background" or "aftershocks", several techniques exist in the literature that are used to assign these tags to the earthquakes. These methods creatively use the robust empirical laws such as Omori law, Gutenberg Richter law, productivity law and so on to identify the aftershocks in the earthquake catalogues, which have been known to adhere to these laws along with the assumption that the "background" earthquakes follow a Poisson process.

In this study, we have only considered three out of many declustering methods that exist in the literature. The first one is based on the SVETAS model, while the other two are based on the method proposed by Reasenberg [1985] and Zaliapin et al. [2008], respectively.

## 3.2.2.1 Smoothed seismicity forecast based on the background component of the SVETAS model (D-SVETAS)

In the case of the SVETAS-based declustering, we obtain the independence probabilities corresponding to each earthquake, which we have defined earlier to be the probability that the earthquake has not been triggered. These independence probabilities are not binary values (0 or 1) but can assume any value between 0 and 1. Smoothing these independence probabilities allows us to obtain the smoothed number of background earthquakes in each of the spatial bin during the training period. The smoothing is performed in the same manner as in Section 3.2.1.3. From this, we construct the forecast of the total seismicity rate during the testing period by first normalising the spatial density of the background earthquakes such that it integrates to 1 over the predefined testing polygon and then by upscaling the spatial PDF by a factor $\frac{T_{test}}{T_{train}} N(\geq M_t)$, where $N(\geq M_t)$ is the number of earthquakes with magnitude larger than the predefined magnitude threshold, $M_t$, of the testing catalogue that have been observed during the training period within the testing polygon. This upscaling is a necessary inconsistency as, in the original state, the smoothed seismicity model would only forecast the rate of background earthquakes during the testing period, which is just a part of seismicity that is going to occur during that period.

## 3.2.2.2 Smoothed seismicity forecast based on Reasenberg's declustering algorithm (D-HKJ)

This method was introduced by Reasenberg [1985] and is based on the previous work of Savage [1972]. In this algorithm, an earthquake that occurs within the spatiotemporal interaction zone of an earthquake cluster composed of preceding earthquakes, can be linked to the cluster. The interaction zone of any given cluster is defined by the time, location and magnitude $\{t_{max}^k, x_{max}^k, y_{max}^k, M_{max}^k\}$ and $\{t_{last}^k, x_{last}^k, y_{last}^k, M_{last}^k\}$ of the largest and last event of the cluster with index k, respectively. Any earthquake i following the last earthquake of the cluster can be potentially associated to the cluster k if its distance from the last and largest earthquake of the cluster is smaller than $r_{fact} 10^{A \times M_{last}^k - B}$ or $10^{A \times M_{max}^k - B}$, respectively. Different authors have suggested different values of the parameters A and B. In this article, we have fixed the value of A and B to 0.5 and 2, respectively [Helmstetter et al.,

2007; Wells and Coppersmith, 1994]. $r_{fact}$ is an arbitrary positive multiplicative factor, which is used to upscale (or downscale) the interaction zone of the last event of any cluster. For the $i^{th}$ earthquake to be definitely associated to the $k^{th}$ cluster, the time elapsed between the $i^{th}$ event and the last event of the cluster ($t_i - t_{last}^k$) should be smaller than $\tau$, where $\tau$ is defined as a "look ahead time" and is equal to $\left\{\log\left[\frac{1}{(1-P)}\right](t_{last}^k - t_{max}^k)\right\} 10^{-\frac{2(M_{max}^k - (M_c + x_k \times M_{max}^k) - 1)}{3}}$. The definition of the look ahead time is derived using the Omori law with exponent 1, a pre-factor $10^{-\frac{2(M_{max}^k - (M_c + x_k \times M_{max}^k) - 1)}{3}}$ and a predefined confidence level $P$. The pre-factor effectively accounts for the productivity of the cluster depending on the magnitude of the largest event in the cluster and assumes a completeness level of $M_c + x_k \times M_{max}^k$, which is increased over the base magnitude of completeness $M_c$ (=2.95 in our case) by an amount $x_k \times M_{max}^k$.

Since the value of $\tau$ can vary between 0 and $\infty$, the upper and lower bounds ($\tau_{max}$ and $\tau_{min}$) must be specified as well.

If an event is associated with a previously clustered event, it becomes a member of the existing cluster. Furthermore, when two events belonging to different clusters are associated, the respective clusters are redefined as a single cluster.

The declustering method has 5 free parameters: $r_{fact}$, $x_k$, $P$, $\tau_{min}$, and $\tau_{max}$. In this study, we have set these parameters to 8, 0.5, 0.95, 1 day and 5 days, respectively. These values have been prescribed in [Helmstetter et al., 2007]. However, we also explore the sensitivity of the forecasting results to these parameters in Text S2. The outcome of applying this declustering procedure on the training catalogues with the fixed set of parameters is the "hard" classification of earthquakes into two groups: independent earthquakes and dependent earthquakes with independence probabilities 1 and 0 respectively. Having obtained the independence probabilities for the earthquakes in the training catalogue, the forecast for the testing catalogue is constructed identically as in Section 3.2.2.1.

**3.2.2.3 Smoothed seismicity forecast based on Zaliapin declustering algorithm (D-ZB)**

Zaliapin et al. [2008], following the work of Baiesi and Paczuski [2004], proposed a simple declustering algorithm based on a rescaled space-time distance between earthquakes. This declustering algorithm is composed of the following steps:

1. We identify the nearest neighbour earthquake, with index $i^*$, to the $j^{th}$ earthquake in the catalogue. To find the nearest neighbour earthquake, we first compute the space-time distance, $n_{ij}$, between the $j^{th}$ earthquake and all the preceding earthquakes. Following Baiesi and Paczuski [2004], $n_{ij}$ is defined by

$$n_{ij} = t_{ij}^{\theta} r_{ij}^{d_f} 10^{-bm_i} \qquad (7)$$

In the above equation, $i$ (= 1 to $j-1$) is the index of each of the earthquakes preceding the $j^{th}$ earthquake; $i^*$ corresponds to the index $i$ for which $n_{ij}$ is minimum; $t_{ij} = t_j - t_i$; $r_{ij} = \sqrt{(x_j - x_i)^2 + (y_j - y_i)^2}$; $\{t_i, x_i, y_i, m_i\}$ is time, location and magnitude of the $i^{th}$ earthquake; $\theta$ is the exponent of the Omori law; $d_f$ is the fractal dimension of earthquake epicentres and $b$ is the exponent of the GR law. Note that we denote the minimum space-time distance between events $j$ and $i^*$ as $n_{i^*j}$.

2. We repeat step 1 for $j = 2$ to $N$, where $N$ is the total number of earthquakes in the catalogue, to obtain the individual $n_{i^*j}$'s for all earthquakes in the catalogue except the first one. The first earthquake is by default considered as a background earthquake.

3. To further classify events $j=2$ to $N$ as background or aftershock, we then perform a 1D Gaussian Mixture modelling on the $\log_{10} n_{i^*j}$ data. We apply a standard Gaussian mixture modelling to these N-1 data points, assuming that the data can be described by a mixture of two Gaussian components (one corresponding to the background earthquakes and the second one corresponding to the aftershocks), with means and standard deviations $\{\mu_1, \mu_2\}$ and $\{\sigma_1, \sigma_2\}$, respectively, and mixing coefficients $\{\pi_1, \pi_2\}$. We then use a standard EM algorithm to estimate $\{\mu_1, \mu_2\}$, $\{\sigma_1, \sigma_2\}$ and $\{\pi_1, \pi_2\}$, and compute the responsibilities $\gamma_{1,j}$ and $\gamma_{2,j}$ that each component of the Gaussian mixture holds for the j$^{th}$ earthquake. Note that $\gamma_{1,j} + \gamma_{2,j} = 1$. In other words, $\gamma_{1,j}$ represents the probability for the j$^{th}$ earthquake to be part of the Gaussian cluster with index 1, which in our case represents the cluster of background earthquakes. Note that we use the property that the mean space-time distance of the background component is larger than that of the clustered component to automatically assign the label "background" to the group of earthquakes associated with one of the two identified clusters. As a result, we obtain the independence probability corresponding to the j$^{th}$ earthquake, $IP_j = \gamma_{1,j}$.

Note that the Zaliapin declustering algorithm is characterised by three parameters θ, $d_f$, and b. Zaliapin et al. [2013] showed that the declustering method is rather robust in identifying background earthquakes in synthetic ETAS catalogues concerning the specific choice of values for these parameters. In this study, we have set these parameters to typical values: θ = 1, $d_f$ = 1.6 and b = 1 respectively [Zaliapin et al., 2013]. Nevertheless, we also explore the sensitivity of the forecasting results to these parameters in Text S2.

Having obtained the independence probabilities for the earthquakes in the training catalogue, the forecast for the testing catalogue is constructed identically as in Section 3.2.2.1.

**3.2.3 Simple Smoothed Seismicity (SSS) model and a model based on Strain rate data (BL)**

In the simple smoothed seismicity approach [Zechar and Jordan, 2008], no distinction is made between a background earthquake and an aftershock. All earthquakes are treated equal and independent of each other. All earthquakes in the training catalogue are smoothed using isotropic Gaussian kernels with a nearest-neighbour dependent bandwidth. Doing so, we obtain the smoothed number of earthquakes in each of the testing cells, which is then scaled as in Section 3.2.2.1 to get the forecast of the SSS model for the testing period.

We do not construct the Bird and Liu's (BL) model forecast in this study but merely use the rates of $M \geq 5.66$ earthquakes in [Bird and Liu, 2007] for the study region and scale it appropriately (as in Section 3.2.2.1) to obtain forecasts for any testing period.

In Figure 2, we show the forecast for M ≥ 4.95 earthquakes for the year 2011 based on these six models as a demonstration. Furthermore, in Text S1, we discuss the differences between the forecasts obtained from these models.

**3.3 Forecast Evaluation**

To evaluate the forecasting performance of the different models, we use the metric of sample information gain per earthquake as defined by Rhoades et al. [2011]. The information gain that model A provides over model B for the $i^{th}$ earthquake during the testing period is given by: $IG_i = \log \lambda_{jk}^A - \log \lambda_{jk}^B - \frac{\hat{N}_A - \hat{N}_B}{N}$, where $\lambda_{jk}^A$ and $\lambda_{jk}^B$ respectively are the seismicity rates forecasted by models A and B in the $j^{th}$ testing grid cell and $k^{th}$ magnitude bin in which the $i^{th}$ earthquake occurred during the testing period; $\hat{N}_A$ and $\hat{N}_B$ are respectively the total number of earthquakes forecasted by models A and B; N is the number of earthquakes that are actually observed during the testing period. Since all models in this study assume the same magnitude distribution for the forecasted earthquakes (i.e. a GR law with b=1), the $IG_i$ is effectively equivalent to $IG_i = \log \lambda_j^A - \log \lambda_j^B - \frac{\hat{N}_A - \hat{N}_B}{N}$.

Having obtained the pairwise information gain of one model over another, we also evaluate if the mean and median of the obtained sample information gain is significantly different from 0 using a pairwise t-test (T-test) and a sign rank test (W-test), respectively [Rhoades et al., 2011].

A model is deemed significantly better if the p-value obtained from these tests is smaller than the standard significance threshold of 0.01.

**4 Results and Discussion**

**4.1 Overall performance of the models in the 6 experiments**

In Figure 3a, the boxplots show the estimates of the mean information gain per earthquake (MIGPE) and its confidence intervals (25-75% CI using colored boxes, and 2.5-97.5% CI using dashed bars) as a function of different model types. It is important to note that, in figure 3a we evaluate the information gain of all the models with respect to a spatially and temporally homogeneous Poisson process (STHPP hereafter). The STHPP model forecasts the same number of earthquakes in each of the testing grid cell, which is equal to $\frac{N(\geq M_t)}{N_{grid}} \times \frac{T_{test}}{T_{train}}$, where $N(\geq M_t)$ is the number of earthquakes with magnitude larger than the predefined magnitude threshold, $M_t$, of the testing catalog that have been observed during the training period within the testing polygon; $N_{grid}$ is the number of 0.1° X0.1° grid cells that divide the testing polygon; $T_{test}$ and $T_{train}$, respectively, are the duration of the testing and the training periods. As for the other models, we also assume that the magnitudes of the earthquakes forecasted by the STHPP model follow the Gutenberg Richter law with a spatially invariant b-value equal to 1. In the complementary Figure 3b, we choose D-HKJ model instead of STHPP as the reference and compute the information gain of all the models with respect to it. As a result, the D-HKJ model is missing from the model list in Figure 3b.

In both the figures, different colors encode the estimates corresponding to the 6 experiments described above with varying settings in terms of minimum magnitude threshold and time duration of the target catalog. The line y=0 marks the line of no information gain, which if we were showing probability gains would be equivalent to y=1.

By pairwise comparison of the MIGPE obtained for all possible pairs of models, we find that SVETAS model consistently obtains higher information gain compared to all the competing models, indicating its superiority. This observation is robust with respect to whether we choose STHPP or D-HKJ model as the reference model.

It is also clear from Figure 3b that GETAS model obtains consistent positive, although marginal information gain over D-HKJ model making it possibly the second-best model in the set of models considered in this paper.

D-ZB model obtains nearly no information gain over the D-HKJ model in Experiments 1 and 4. In all the other experiments D-ZB consistently loses information with respect to D-HKJ model. The similarity in the forecasting prowess of the two models is not surprising considering the similarity between the forecasted spatial density of the two models (see for instance Figure 2) which the two models have despite apparently different declustering ideologies.

We also notice from Figure 3b that the SSS model consistently underperforms with respect to D-HKJ model in all the experiments. This observation is indeed consistent with the findings of Werner et al. [2011] who also found that declustering the training catalog using the Reasenberg algorithm prior to obtaining a smoothed seismicity forecast (as in case of D-HKJ model) is a better strategy in forecasting the location of future earthquake than directly obtaining a smoothed seismicity forecast (as in case of SSS model).

The D-SVETAS model, which is based on the declustered component of the SVETAS model, consistently underperforms with respect to D-HKJ model. In fact, it tends to underperform even with respect to SSS model as well at least in terms of MIGPE shown in Figure 3. This raises a question on the general efficacy of the working hypothesis that location of past background earthquakes reveal the probable location of future seismicity. It seems that the specificity of declustering algorithms tend to play an important role in forecasting performance of the resulting SSMs. Indeed, these specificities, which include the general idea as well as the value of the declustering parameters, control the extent to which the training catalog is declustered and hence the resulting forecasts for the testing periods. In case of D-SVETAS and D-ZB models the parameters that control the declustering can be obtained using a self-consistent framework. However, for the Reasenberg's declustering algorithm, on which the D-HKJ model is based, no such self-consistent framework exists and we had to rely on some "standard" parameter combinations proposed by Helmstetter et al. [2007].

Based on the overall forecasting performance of the models shown in Figure 3, one can argue that among the three declustering methods used in this study, Reasenberg's declustering algorithm is the best in distinguishing between aftershocks and background earthquakes, as it best forecasts location of all future earthquake. However, such a conclusion is inherently flawed. In none of the experiments that we have conducted in this study, we try to forecast the location of future background earthquakes. In fact, such an experiment would be inherently biased if not impossible. This is because nature does not give us access to the information of whether an earthquake is a background earthquake or not. Such definitions are only anthropogenic constructions. So, if in a hypothetical forecasting experiment we define earthquakes during the testing period as background and triggered using certain declustering method and evaluate the performance of all models based on the limited set of background earthquakes, we would inherently favor the model that is the most consistent with the declustering method used for the definition itself. This would then render the whole forecasting experiment useless.

Finally, we notice that the BL model, which utilizes the many geologically relevant information such as strain rate, location and orientation of faults and so on tends to underperform relative to all purely statistical models. Such gross underperformance possibly indicates that at the time scales of forecasts considered in this study, those geological information alone cannot lead to forecasts superior than the purely statistical models. Either these models should be reevaluated on significantly longer forecasting horizons are they should be combined with their statistical counterparts to yield ensemble forecasts. In fact, Rhoades et al. [2014] showed that for the same testing region used in this study, a multiplicative hybrid of D-HKJ model and BL model yielded better forecast than the two models alone, during the 5 year long testing period [2006, 2010] and for the target magnitude threshold of 4.95.

## 4.2 Is the performance of models significantly different from each other?

To assess if one model performs significantly better than any other in the six experiments, we carried out pairwise comparisons of the model information gains using T and W tests [Rhoades et al., 2011, also see Section 3.3]. The advantage of these pairwise comparisons is that it allows each model to serve as a reference relative to all the other models, thus leading to a direct comparison between the two rather than an indirect comparison relative to some common null model.

In Figure 4 (top panels), we show the *p*-values obtained from pairwise comparisons of the information gains of the seven models in all the six experiments using the T-test. A similar plot, but with the *p*-values obtained with the W test, is shown in the bottom panels of Figure 4. We treat model A to be significantly better than model B if the *p*-value obtained from the test, which is reported in row A and column B, is smaller than 0.01 (-2 in log10 scale). All the cases in each experiment, where we were able to find significant evidence in favour of model A over B, have been highlighted using red rectangles. Note that, the p-values that were smaller than the computer precision have been marked as -inf.

Treating each pairwise comparison as a game and a win declared only if one model wins with acceptable statistical significance, we find that SVETAS, GETAS, D-SVETAS, D-HKJ, D-ZB, SSS and BL models win 36, 23, 4, 19, 16, 8 and 0 games, respectively, out of 36 games that each model played. The results from the W-test shown in Figure 4 (bottom panels) are similar with SVETAS, GETAS, D-SVETAS, D-HKJ, D-ZB, SSS and BL models winning 36, 23, 7, 16, 18, 5 and 0 games, respectively. Combining the results from the two significance tests, we find the ranking: SVETAS > GETAS > D-HKJ ≅ D-ZB > SSS ≅ D-SVETAS > BL. Note that, this ranking is also almost consistent with the boxplot of MIGPE values shown in Figure 3.

We also note that the superiority of SVETAS is overwhelming, in the sense that the largest p-values, which results from the direct comparison on SVETAS model and the GETAS model in Experiment 4 (testing period being 1 year and $M_t = 4.95$), of all the pairwise comparisons in which SVETAS is involved are always found smaller than $10^{-4.9}$. Thus, at any standard confidence level of 95% or 99%, the null hypothesis is rejected. Even very skeptical readers, who would require a confidence level of 99.998%, should reject the null when presented with our p-value of $10^{-4.9}$. In comparison, for instance, the second-best performing model, GETAS, sometimes wins over other models with just p=0.01, which will allow one to reject the null hypothesis only when deciding on a 95% or marginally on a 99% confidence level.

We also notice that the strength of evidence in favour of the winning models increases with the decrease in minimum magnitude threshold of the testing catalogue, which can be explained by the increasing number of earthquakes in the testing catalogue.

### 4.3 Time Evolution of the Information Gain

Figure 5 shows the time series of the cumulative information gain (CIG) for the seven models, over the spatially and temporally homogeneous Poisson process (STHPP). A complementary plot showing the time series of the average information gain per earthquake (AIGPE) over STHPP for all models is shown in Figure 6 where we have obtained the AIGPE values in sliding time windows that contain 20 earthquakes for Experiments 1 and 4, 200 earthquakes for Experiments 2 and 5 and 2000 earthquakes for Experiments 3 and 6. The time series is obtained by plotting the AIGPE values obtained for each sliding window against the time of the last earthquake in the window.

From the two sets of time series, we can observe that the SVETAS model consistently outperforms all the other competing models in all the experiments. Its time series showing its cumulative information gain and average information gain per earthquake are growing faster and above from those of other models and stays on top at nearly all times during the entire testing period. This confirms that the SVETAS model is the best among all the seven models, and its over-performance is not due to some lucky period but is consistent over the entire testing period.

We also find that even though the GETAS model ends in the second place in all experiments, its cumulative information gain and average information gain per earthquake time series is not always clearly separated from those of D-HKJ and D-ZB models. This is especially true for experiments 1, 2, 3 and 4. For experiments 5 and 6, the GETAS model is a definite runner-up. These observations seem to indicate that the performance of the GETAS model relative to other competing models tends to improve when time duration and minimum magnitude thresholds of the testing catalogues get smaller, allowing it to forecast short-term clustered seismicity in the testing data based on the relatively recent earthquakes in the training catalogue. D-HKJ and D-ZB models perform nearly identically in all experiments, which is consistent with the fact that the forecasts of the two models are highly correlated with each other. The cumulative information gain curve of the D-SVETAS and BL models reveal that they are the weakest in all the experiments. However, these two models, along with the SVETAS model, get a significant boost in performance in the year 2010, when the $M_w$ 7.2 El-Major Cucapah earthquake occurred. All the other models exhibit a deteriorated performance following the occurrence of this event. All models tend to perform badly in year 2014 (for magnitude cutoffs of 2.95 and 3.95) due to the occurrence of an intense swarm activity in northwest Nevada in November of that year. All models consistently under-forecasted in the swarm region. As the earthquakes that occurred in this swarm sequence were mostly smaller than 4.95, no effect can be seen when using a magnitude threshold of 4.95.

**4.4 Consistency of observed and forecasted number of earthquakes**

In Figure 7, we show the consistency between the observed (black crosses) and the number of forecasted earthquakes (colored squares) by all the models for the six experiments during all the testing periods. While the SVETAS and GETAS models can forecast the total seismicity rate in any testing period, all other models (grouped under the class Rest) use the observed number of earthquakes in the training period (normalised by the duration of testing period) to estimate the forecast for the testing period. In order to test the consistency of the observed numbers with forecasts for a given testing period, we then assume (in coherence with the "N-test" proposed by CSEP) that the observed numbers can deviate from forecasts within limits (95% quantiles) set by a Poisson PDF with mean rate being equal to the forecasted number. These limits of deviations are shown using the colored error bars for the three model types. If the deviations exceed these set limits, the forecast can be said to be inconsistent with the observation.

We find that the models tend to be mostly inconsistent for the magnitude threshold of 2.95, but the consistency improves with increasing magnitude threshold. For the target magnitude threshold of 4.95, the consistency of the forecasts with the observed number is a rule rather than an exception. However, this emerging rule should not be confused for a "true" consistency. Instead, it only reflects that there are not enough events ($M \geq 4.95$) in any given testing period to allow one to conclude about the consistency of the rates forecasted by different models with the observed number of earthquakes under an arbitrarily imposed Poissonian assumption.

Several factors contribute to the inconsistency of the forecasted and observed number of earthquakes. First, the test assumes that the distribution of the number of earthquakes within a given testing period is Poissonian. As a result of this assumption, the range of expected deviation for a given forecast is exceptionally narrow. Several authors [Werner and Sornette, 2008; Werner et al., 2011; Jackson and Kagan, 1999; Saichev and Sornette, 2006] have argued that the distribution of the number of earthquakes (in the real catalogue) within a time bin of given duration is, in fact, better explained by distributions with tails thicker than a Poisson distribution. For instance, Werner et al. [2011] showed that the empirical distribution of the number of earthquakes with $M \geq 4.95$ in the ANSS catalog from 1932 to 2008 within the CSEP testing region surrounding California is better described by a negative binomial distribution compared to a Poissonian distribution. Saichev and Sornette [2006] have shown that the number of earthquakes within finite space-time windows in California can be represented by asymptotic power law tails, which is also consistent with the predictions of ETAS model. Werner and Sornette [2008] studied the impact of magnitude uncertainties on rate estimates in clustering models, on their forecasts and on their evaluation by CSEP's consistency tests. Given that magnitude uncertainties are more heavy-tailed than a Gaussian, they found that the Poissonian assumption of the consistency tests is inadequate for short-term forecast evaluations.

Another possible cause of the inconsistency of the observed and forecasted number of earthquakes is the non-stationarity of the seismicity rate. Not only there are unknowable short-term fluctuations in the seismicity rate following the large earthquakes (such as in years 1999 and 2010), there could also exist non-stationarity at longer timescale. In Figure 8 (upper panel), we show the time series of the cumulative number of $M \geq 2.95$ "background" earthquakes within the CSEP collection polygon surrounding California. The four lower panels in the figure show the time series of the number of background earthquakes within a 1-year time window and a sliding duration of 30 days. These background earthquakes are obtained using the four declustering approaches considered in this study: D-HKJ, D-ZB, SVETAS and GETAS. As a result of declustering, the major short term fluctuations in seismicity rate are removed. The remaining events are expected to follow a Poisson process with no systematic time structure in the seismicity rate. However, we find that, despite the disagreement between the total expected number of background earthquakes, all the declustering methods consistently agree in one respect: the rate of background earthquakes in California is non stationary and seems to systematically decrease during the study period. It is clear that, if one uses the training catalog (say from the 1981 to 1999), one is bound to overestimate (significantly) the background seismicity rate for the testing period that follows. This overestimation would not be nullified if one takes account of the aftershocks, as the total number of aftershocks is only a multiplicative factor of the number of background earthquakes that are expected to occur (at least in the hypothetical ETAS universe). Indeed, this hypothesis is coherent with our observations in Figure 7 (see for instance Row 1, Column 1) that all models tend to systematically over-forecast the number of earthquakes during the typical testing periods (those which are not influenced by the intermittent burst of seismicity following a very large event).

## 5 Conclusion

In all the experiments conducted in this study, the ETAS model with spatially variable parameters (SVETAS) comes out as the statistically significant winner over other smoothed seismicity-based models and is followed by the ETAS model with a global set of parameters (GETAS) in most of the experiments. In none of the experiments, a smoothed model using only the location of past background earthquakes could outperform the two ETAS based models, which forecast the rate of aftershocks of all generations. Indeed, our findings highlight the importance of forecasting the rates of future aftershocks, contrarily to the existing paradigm that the background seismicity alone can lead to a superior smoothed seismicity forecast. Our results also provide evidence in favor of extending the applications of ETAS based models (SVETAS in particular) from daily/monthly forecasting horizons to longer forecasting horizons of 1-5 years, and possibly to even longer forecasting horizons, which are relevant for seismic hazard models such as UCERF3 [Field et al., 2015].

Among the two ETAS models themselves, accounting for the spatial variation of the parameters leads to a statistically significant improvement in the forecasting performance of the models and thus provides more credibility to the spatial variation in parameters reported for the SVETAS model.

Finally, we believe that the "true" performance of a model can only be revealed in a completely prospective experiment. As a result, we further aim to participate in the forecasting competitions organized by CSEP to further increase our confidence in the robustness and reliability of our results.

**Acknowledgement**

The dataset used in this study can be obtained from the website.

http://www.quake.geo.berkeley.edu/anss/catalog-search.html.

S.N. wishes to thank Y. Kamer for participating in discussions during the development of the work; M.J Werner, A. Helmstetter, D. Jackson and an anonymous reviewer for critical suggestions and discussions, which led to significant improvement of the manuscript.

*Table 1: Definition of important acronyms used in the manuscript*

| Acronyms | Description |
|---|---|
| ETAS | Epidemic Type Aftershock Sequence Model; see equation 1 and 2 |
| GETAS | ETAS model with spatially homogenous parameters; see section 3.2.1.1 |
| SVETAS | ETAS model with spatially variable parameters; see section 3.2.1.2 |
| SSM | A smoothed seismicity model |
| RELM | Regional earthquake likelihood model; see Schorlemmer and Gerstenberger [2007] |
| CSEP | Collaboratory for the Study of Earthquake Predictability; see Werner et al. [2010] |
| D-HKJ | A smoothed seismicity model obtained by smoothing the location of background earthquakes obtained from Reasenberg's declustering algorithm; see section 3.2.2.2 |
| D-ZB | A smoothed seismicity model obtained by smoothing the location of background earthquakes obtained from Zalliapin and Ben-Zion's declustering algorithm; see section 3.2.2.3 |
| D-SVETAS | A smoothed seismicity model obtained by smoothing the location of background earthquakes obtained from SVETAS model; see section 3.2.2.3 |
| SSS | A simple smoothed seismicity model based on undeclustered data; see section 3.2.3 |
| BL | A model proposed by Bird and Liu, [2007], which is based on strain rate data; see section 3.2.3 |
| IG | Information Gain; see section 3.3 |
| MIGPE | Mean information gain per earthquake; see section 4.1 and figure 3 |
| STHPP | Spatially and temporally homogenous Poisson process, which has been defined in section 4.1 |
| CIG | Cumulative information gain; see section 4.3 |
| AIGPE | Average information gain per earthquake; see section 4.3 |

# Figures

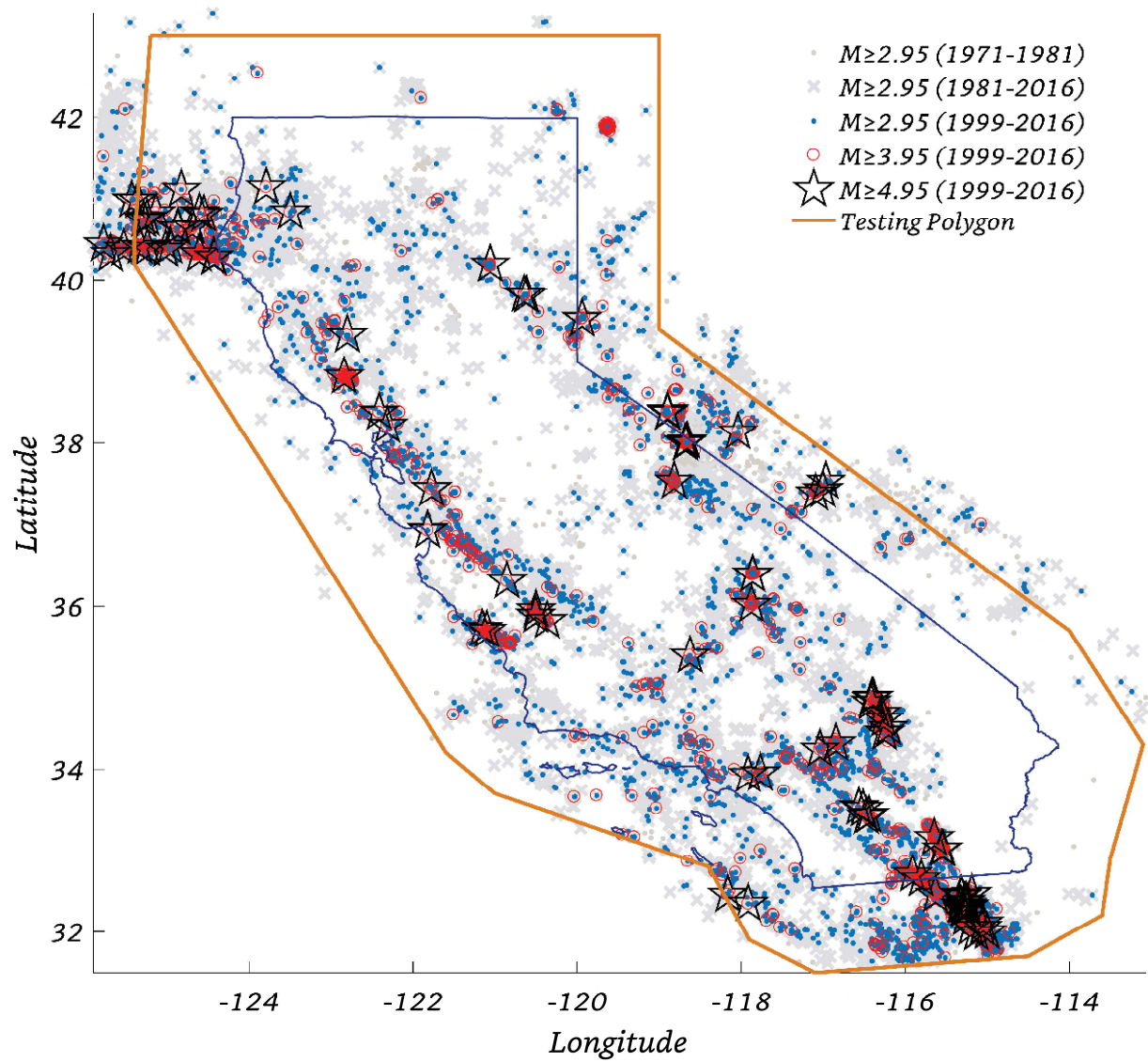

*Figure 1*: *Earthquakes used in this study: grey dots show the locations of auxiliary earthquakes ($M \geq 2.95$; $1971 - 1981$); grey crosses show the locations of all primary earthquakes ($M \geq 2.95$; $1981 - 2016$); blue dots, red circles and black stars show the locations of $M \geq 2.95$, $M \geq 3.95$ and $M \geq 4.95$ earthquakes that occurred during the entire testing period (1999-2016); only earthquakes within the testing polygon are involved in testing.*

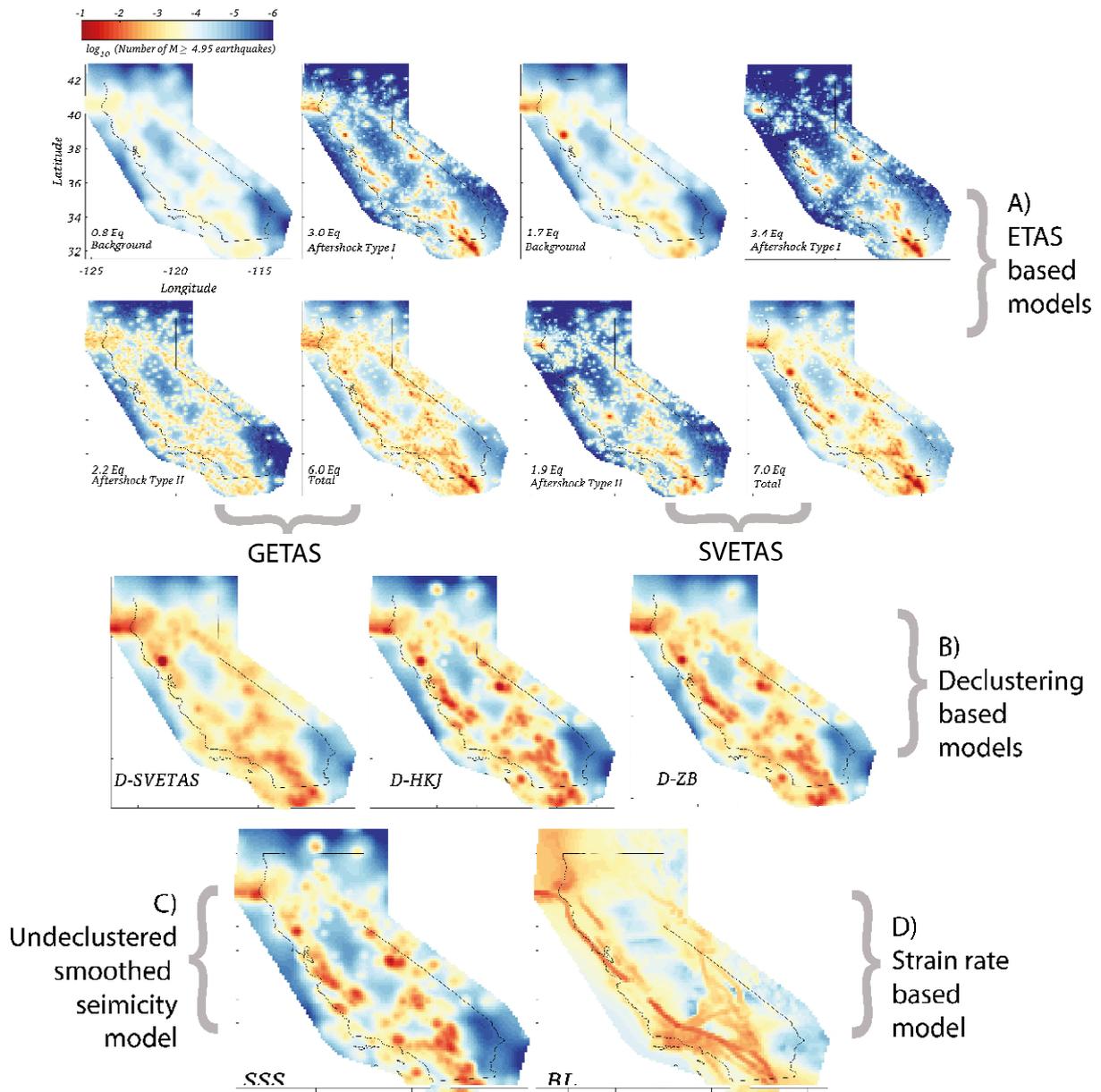

*Figure 2:* Forecast of $M \geq 4.95$ earthquakes for the year 2011 based on six models: GETAS, SVETAS, D-SVETAS, D-HKJ, D-ZB, SSS and BL, investigated in this study; (A) contribution of background earthquakes (Row 1, Column 1 and R1,C3 ), type I aftershocks (R1, C2 and R1,C4) and type II aftershocks (R2, C1 and R2,C1) to the final forecast (R2, C2 and R2, C4) of GETAS and SVETAS models, respectively; (B) declustering based smoothed seismicity models: D-SVETAS, D-HKJ and D-ZB; (C) undeclustered smoothed seismicity model or SSS model; (D) BL or strain rate based model.

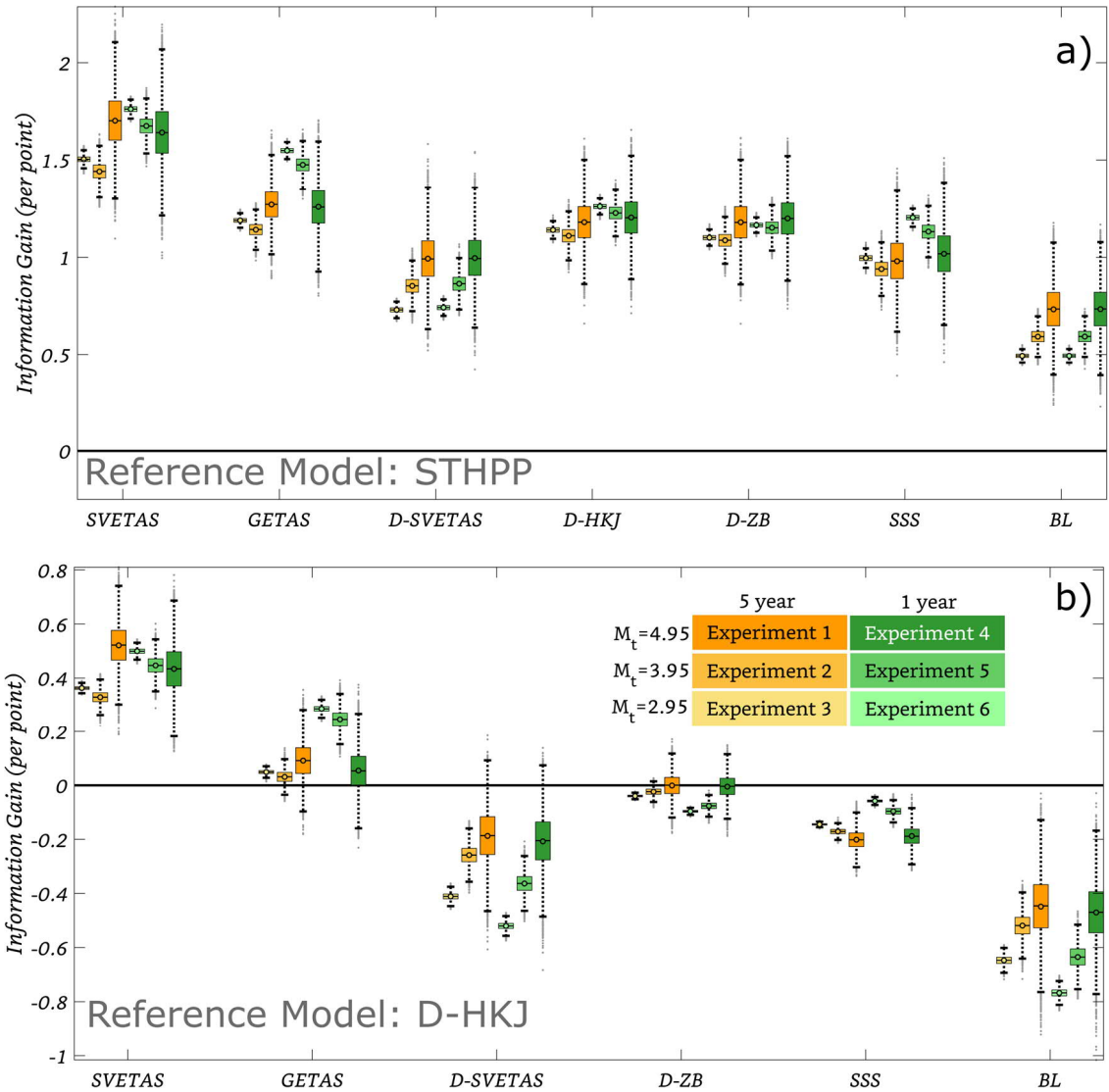

*Figure 3:* Boxplots showing the estimates of the mean information gain per earthquake (MIGPE) and its confidence interval as a function of model type, magnitude threshold of the target catalog, and duration of the testing period; information gain of all competing models has been evaluated with respect to the STHPP and D-HKJ model in panel a and b respectively on each box, the central dashed mark indicates the median, circle indicates the mean, the bottom and top edges of the box indicate the 25 and 75 percent confidence interval, respectively, and the dashed bars indicate the 95% confidence interval; the small dots at the top and bottom of boxes indicate the outlying estimates of the median information gain per earthquake.

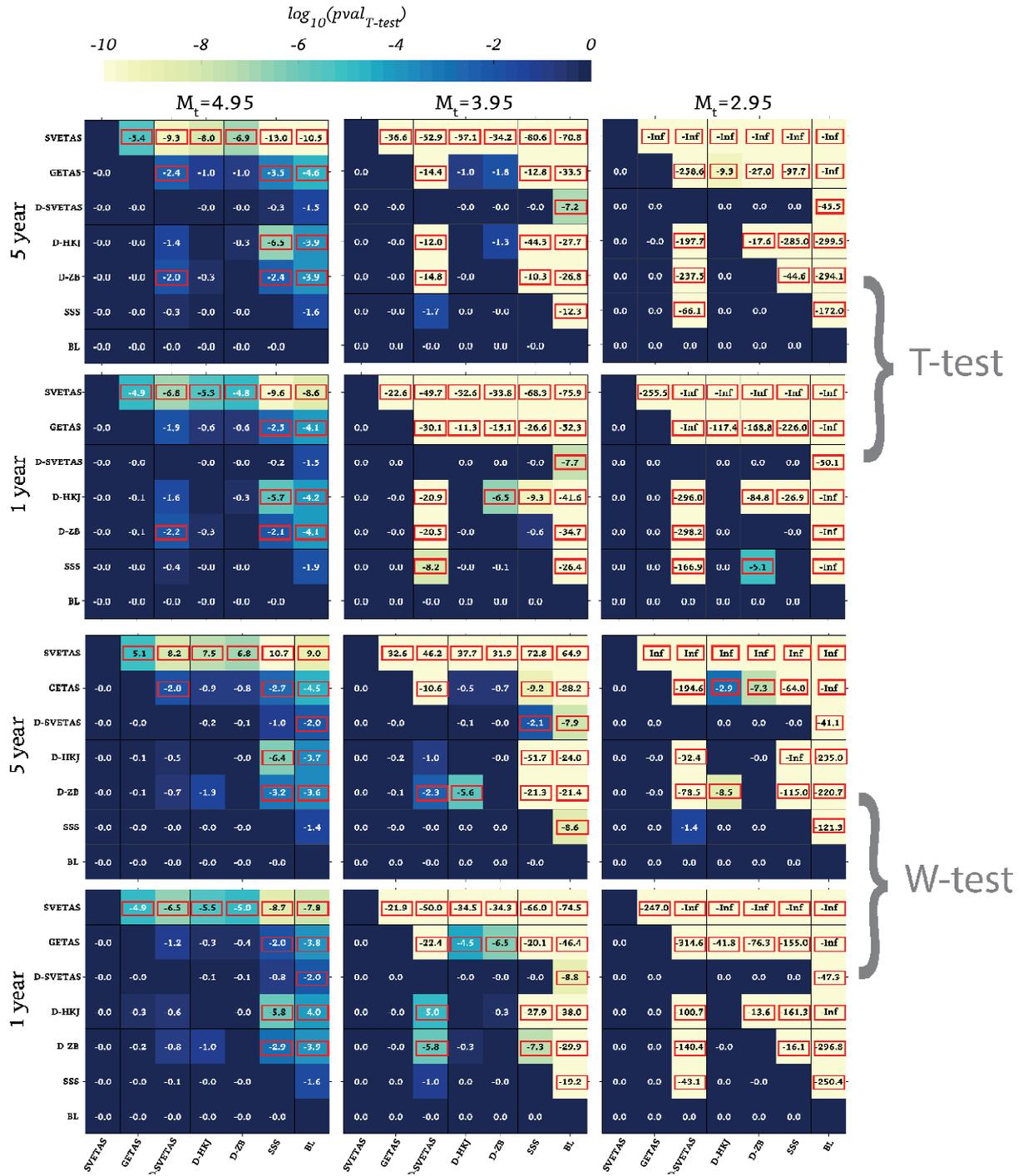

***Figure 4***: *pairwise comparison of information gains of models using T-test and W-test for the 6 experiments; time duration and magnitude threshold of the testing catalogue in each of the six experiments are indicated at the left and top of the figure; the colormap shows the $log_{10} p_{value}$ (also given as numbers in each cell) found from the pairwise comparison of the information gain obtained by the models for all the earthquakes involved in the pseudo prospective testing; values smaller than -2 indicate that there is significant evidence in favor of one model over another: all such values have been highlighted; -inf values occur when the p-values are indistinguishable from 0 at computer precision.*

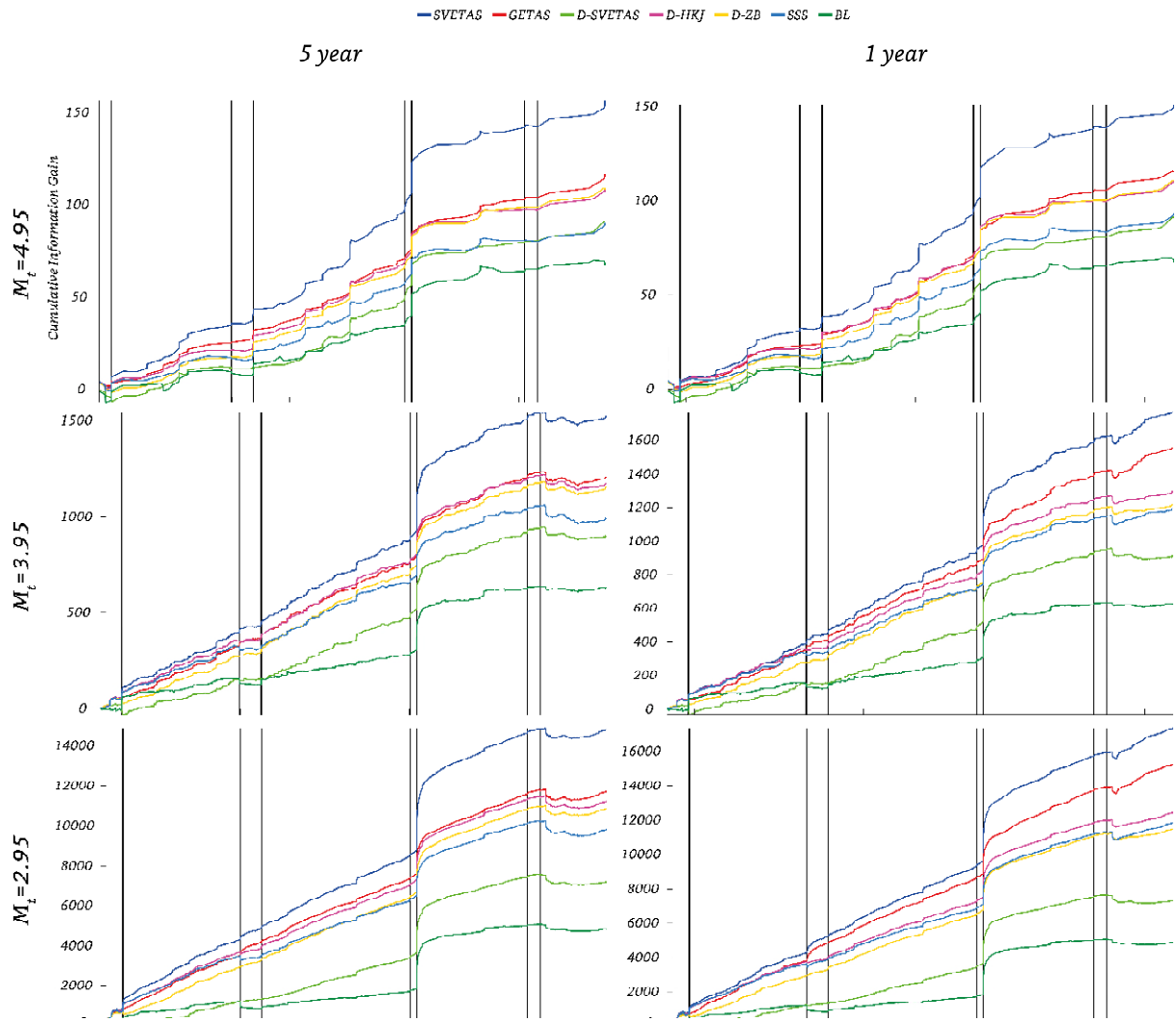

the six experiments conducted in this study; solid black lines mark the time of occurrence of $M \geq 6$ earthquakes during the testing period.

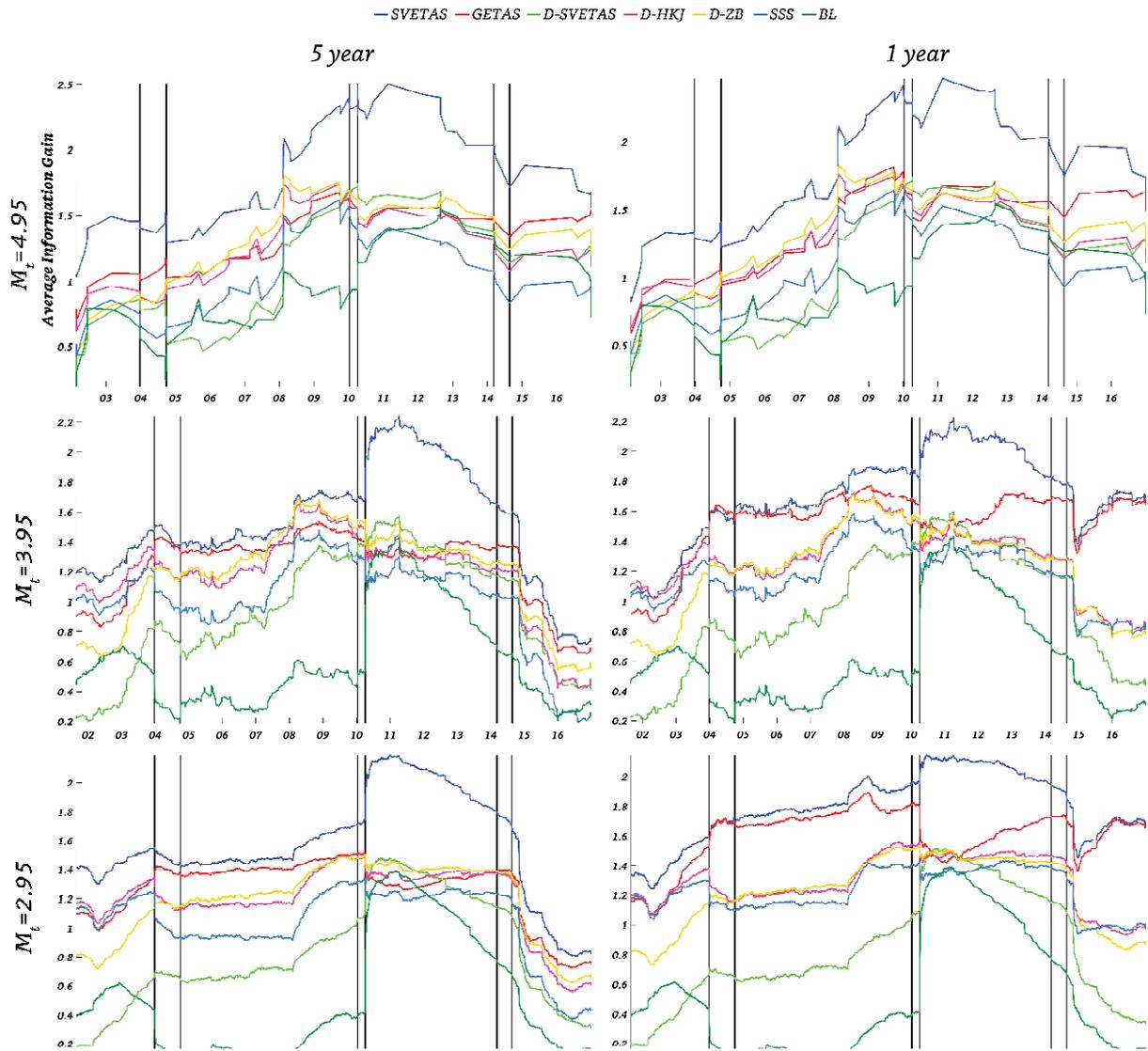

STHPP in the 6 experiments (column 1 corresponds to experiments 1 to 3, and column 2 corresponds to experiments 4 to 6) conducted in this study; solid black lines mark the time of occurrence of $M \geq 6$ earthquakes during the testing period.

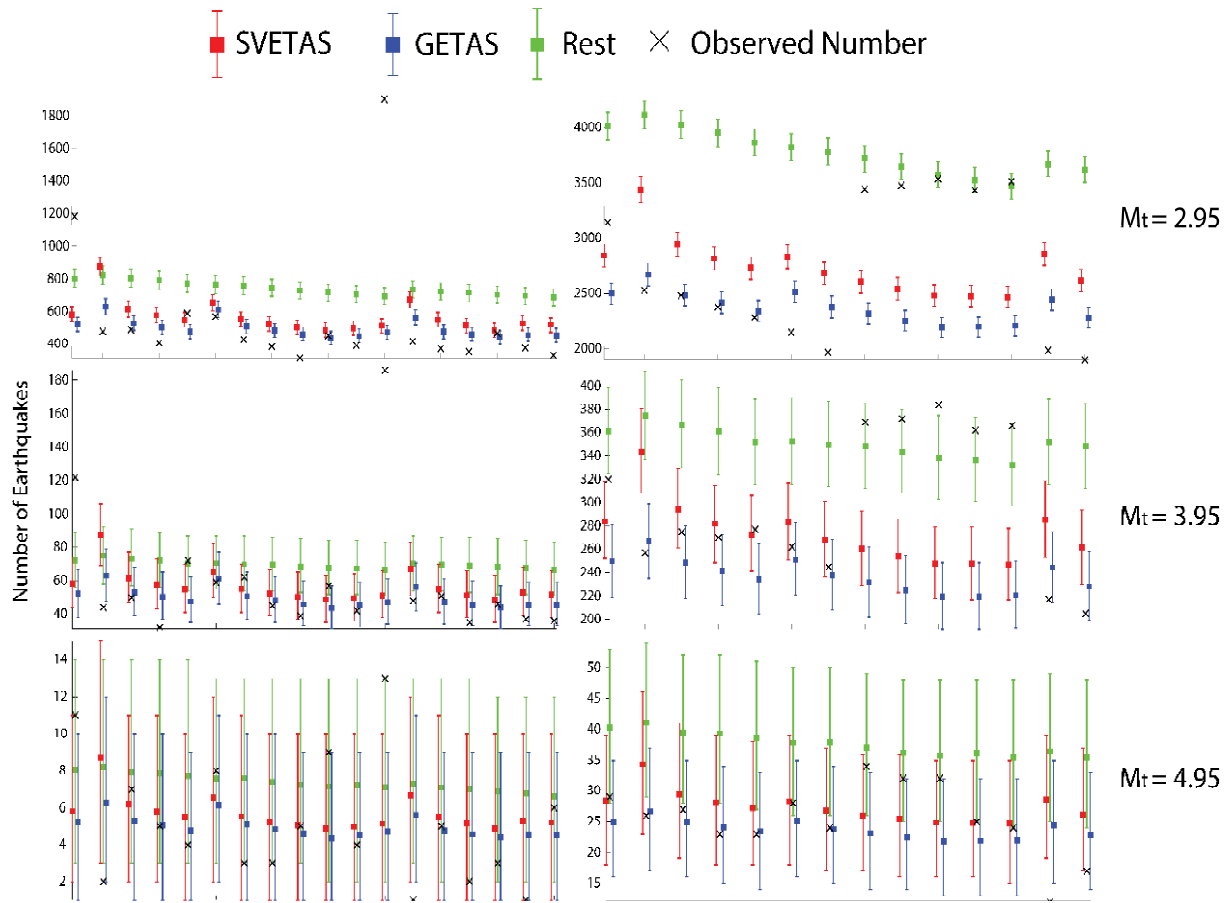

*possible 95% deviation from the forecasted numbers under the Poissonian assumption; Black crosses show the observed number of earthquakes.*

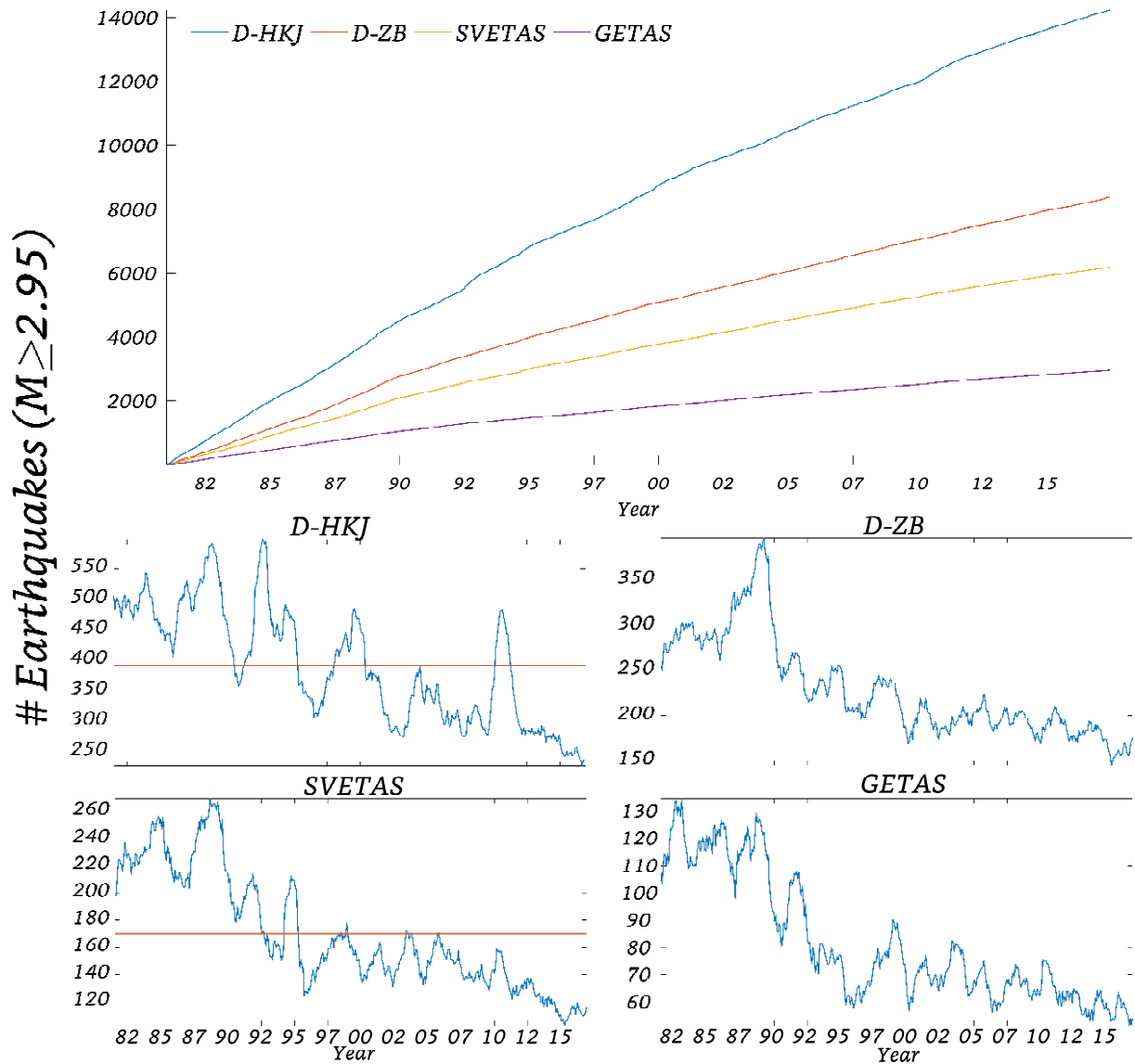

*Figure 8*: *(Upper Panel) Time series of cumulative number of background earthquakes obtained using different declustering methods considered in this study;(Bottom Panels) Time series of number of background earthquakes in a period of 1 year calculated every 1 month for the four declustering methods is shown in blue; Red lines show the average number of background earthquakes during all 1-year periods.*

# Forecasting the rates of future aftershocks of all generations is essential to develop better earthquake forecast models

Shyam Nandan[1], Guy Ouillon[2], Didier Sornette[3], Stefan Wiemer[1]

[1]ETH Zürich, Swiss Seismological Service, Sonneggstrasse 5, 8092 Zürich, Switzerland

[2]Lithophyse, 4 rue de l'Ancien Sénat, 06300 Nice, France

[3]ETH Zürich, Department of Management, Technology and Economics, Scheuchzerstrasse 7, 8092 Zürich, Switzerland

Corresponding author: Shyam Nandan (shyam4iiser@gmail.com)

**Contents of this file**

Text S1 to S2

Figures S1 to S2

**Text S1: How different are the seven smoothed seismicity forecasts from each other?**

In Figure 2, we show the smoothed seismicity forecast of $M \geq 4.95$ earthquakes for the year 2011 based on the seven smoothed considered seismicity models: GETAS, SVETAS, D-SVETAS, D-HKJ, D-ZB, SSS and BL. Note that the forecasts for other magnitude thresholds $3.95$ and $2.95$ can be obtained from these maps by simply multiplying the presented forecasts by a factor $10$ and $100$.

In Figure 2 (A), we show the three essential forecast components: background earthquakes, aftershocks of type I and type II, for the GETAS and SVETAS model. In both models, the background component only explains a meagre ~13% and 24%, respectively, of the total expected seismicity rate. On the other hand, type I and type II aftershocks explain ~50% and 37% for the GETAS model, and ~50% and 26% for the SVETAS model.

It is interesting to note that, while both GETAS and SVETAS models are based on the same "ETAS" principles, obvious differences between them can be identified among the forecasted seismicity patterns for the different components of the two models. For instance, the SVETAS model forecasts much higher background rate and much lower aftershock rates (of both types) in the Geysers geothermal field and offshore Mendocino (barring the triple junction) compared to the GETAS model. This feature is not only present during the chosen testing period but during all the testing periods considered in this study. The aftershock rate forecasted by the SVETAS in the two regions is so exceptionally low that it can be concluded that most of the seismicity in these two regions are the result of exogenous processes. A similar conclusion cannot be reached from the GETAS model. The origin of this difference can be traced to the spatially varying parameters of the SVETAS model. The SVETAS model supports the conclusion that most of the seismicity in the Geysers geothermal field is indeed driven by anthropogenic fluid injection, which is an exogenous process. Furthermore, the higher background seismicity rate in the offshore Mendocino detected by the SVETAS model can be due to either or a combination of the following factors: (i) larger location uncertainty for the offshore events; (ii) larger catalog incompleteness leading to missing parents and orphan earthquakes, which appear as background events; (iii) enhanced external forcing due to some unknown physical mechanism.

In Figure 2 (B), we show the forecasts based on the D-SVETAS, D-HKJ and D-ZB models. We find that the forecasts of the D-HKJ model and D-ZB model are very similar to each other with a correlation coefficient of ~0.95 (i.e. statistically significant) despite apparently different declustering ideologies. On the other hand, the forecast of the D-SVETAS model exhibits a smaller correlation to both models, albeit with a larger correlation with the D-ZB model than with the D-HKJ model (correlation coefficients of ~0.73 and 0.62, respectively).

The extent of declustering achieved by the three models differ. For instance, the 1992 $M_w$ 7.3 Landers and 1999, $M_w$ 7.1 Hector Mine earthquake's aftershock zones are easily observable in the forecasts of the D-HKJ and D-ZB models, while they are almost absent in the forecast of the D-SVETAS model, thus making it the most stringent among the three declustering algorithm. However, it is important to remember that the intensity of declustering itself is not a metric with which one can assess the quality of a declustering algorithm. In fact, the relative performance of the forecasts of the declustering methods can only provide limited evidence in favor of one declustering method over another, as the forecasts are constructed on all the earthquakes in the testing period. These earthquakes not only include the background earthquakes, but also a considerable number of aftershocks, which those declustering methods are designed not to forecast.

Despite the differences in the three declustering methods, all of them consistently forecast high background seismicity (and thus high total seismicity rate) in the region around the Geysers geothermal field and offshore Mendocino. Furthermore, compared to the SVETAS and GETAS models, none of these three models forecast an elevated seismicity rate in the region around the El-Major earthquake.

In Figure 2(C), we show the forecast of $M \geq 4.95$ earthquakes based on a simple smoothed seismicity (SSS) model and a model based on strain rate (BL) in the left and right panels, respectively. While the SSS model bears a high degree of resemblance to the D-HKJ and D-ZB models (correlation coefficients of 0.9 and 0.78, respectively), its resemblance to the D-SVETAS model is rather weak with a correlation coefficient of only 0.4. The systematic decrease of correlation with the SSS model when going from the D-HKJ model to the D-SVETAS model is in line with our previous assertion that the D-SVETAS model is the most stringent of the three declustering algorithms considered in this study.

Finally, the BL model bears the least resemblance to all the smoothed seismicity models considered in this study.

**Text S2: Are Reasenberg's and Zaliapin declustering algorithms with optimal parameters better than the SVETAS model?**

In our tests of the D-HKJ model, we have assumed a fixed set of parameters for the Reasenberg's declustering algorithm based on the work of Helmstetter et al. [2007]. However, other studies such as Schorlemmer and Gerstenberger [2007] have suggested that the parameters $r_{fact}$, $x_k$, $P$, $\tau_{min}$, and $\tau_{max}$ can vary within the range [5-20], [0-1], [0.9,0.99], [0.5, 2.5] and [3,15]. It is thus necessary to investigate the sensitivity of the forecasting potential of the D-HKJ model to the choice of the declustering parameters. For this, we perform 10 regular sampling of each of these parameters and obtain $10^5$ parameter combinations. For each of the combinations, we repeat the experiments 4, 5 and 6. Figure S2 (upper panel) shows the distribution of MIGPE obtained over the $10^5$ parameter combinations for the experiments 4, 5 and 6. For a given parameter combination and experiment type, MIGPE is obtained from the sample information gain of each earthquake in all 1 year-long testing periods between 1999 and 2016. In this figure, we also indicate, using triangular markers, the value of MIGPE corresponding to the declustering parameters used in earlier sections and by Helmstetter et al. [2007]. We find that the values of the declustering parameters used earlier in this study are indeed not optimal and yield MIGPE values of 1.21, 1.33 and 1.34 compared to the MIGPE values of 1.31, 1.40 and 1.44 obtained for the optimal parameter combinations {8.33, 0.67, 0.98, 2.05, 13.67}, {10, 0.11, 0.91, 1.83, 15} and {8.33, 0.22, 0.96, 2.05, 8.33} for the experiments 4, 5 and 6, respectively.

To investigate if the smoothed seismicity model with optimal parameters, D-HKJ*, is significantly more informative than the D-HKJ model, we perform a right tailed W-test. The $\log_{10}$ of p-values for this test are shown in the insets of Figure S2. Indeed, the p-values are much smaller than the standard significance threshold of 0.01 and, hence the D-HKJ* model can be more informative. It is important to note, however, that the D-HKJ* model has 5 free parameters, which should be accounted for when comparing the two models. We do this by means of Wilks' loglikelihood ratio test. Given that the cumulative information gain of the D-HKJ* model over the D-HKJ model is 5.61, 32.93 and 317.39 in the experiments 4, 5 and 6 respectively, the Wilks' statistics is found to be 0.046, $7.38 \times 10^{-13}$ and $6.11 \times 10^{-135}$, respectively. Considering the standard significance threshold of 0.01, the evidence in favor of the D-HKJ* model over the D-HKJ model is quite conspicuous. Finally, even after allowing for parameter optimization of the D-HKJ* model on the testing catalogs, in none of the experiments does the D-HKJ* model surpass the performance of the SVETAS model. On the contrary, the SVETAS model significantly outperforms the D-HKJ* model in each of these three experiments with the p-values of the W-test being much below the standard significance threshold of 0.01 (see insets in Figure S2).

We perform a similar analysis as above for the Zaliapin declustering algorithm, which is characterized by the three set of parameters θ, $d_f$, and b. We allow these parameters to vary within the range (0, 2], (0, 2] and (0, 2]. Within this specified range, we perform 50 regular sampling of each of these parameters and obtain $1.25 \times 10^5$ parameter combinations. Figure S2 (lower panel) shows the distribution of MIGPE for experiments 4, 5 and 6 for these 125,000 parameter combinations. We find that while the value of MIGPE for the optimal D-ZB* model is significantly larger than that of the D-ZB model in all the experiments, the D-ZB* model still significantly underperforms with respect to the un-optimized SVETAS model.

These sensitivity analyses echo our earlier findings that the SVETAS model is a vastly superior smoothed seismicity model compared to D-ZB and D-HKJ, and not even optimization can change this conclusion.

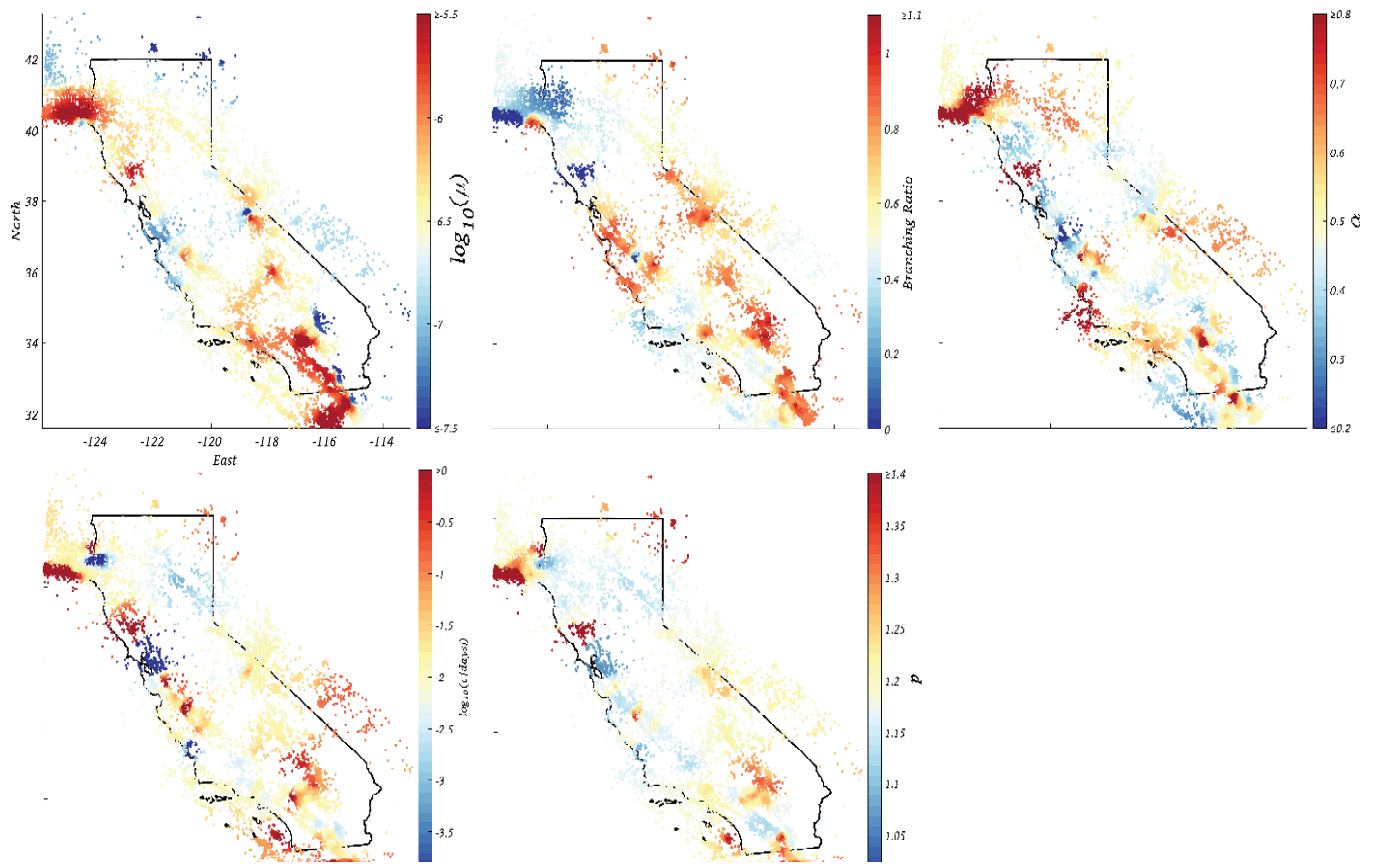

**Figure S1:** Spatial variation of the Background seismicity rate ($\mu$), Productivity parameters: $n$ and $\alpha (= \frac{a}{log(10)})$ and Omori parameters: $c$ and $p (= 1 + \omega)$.

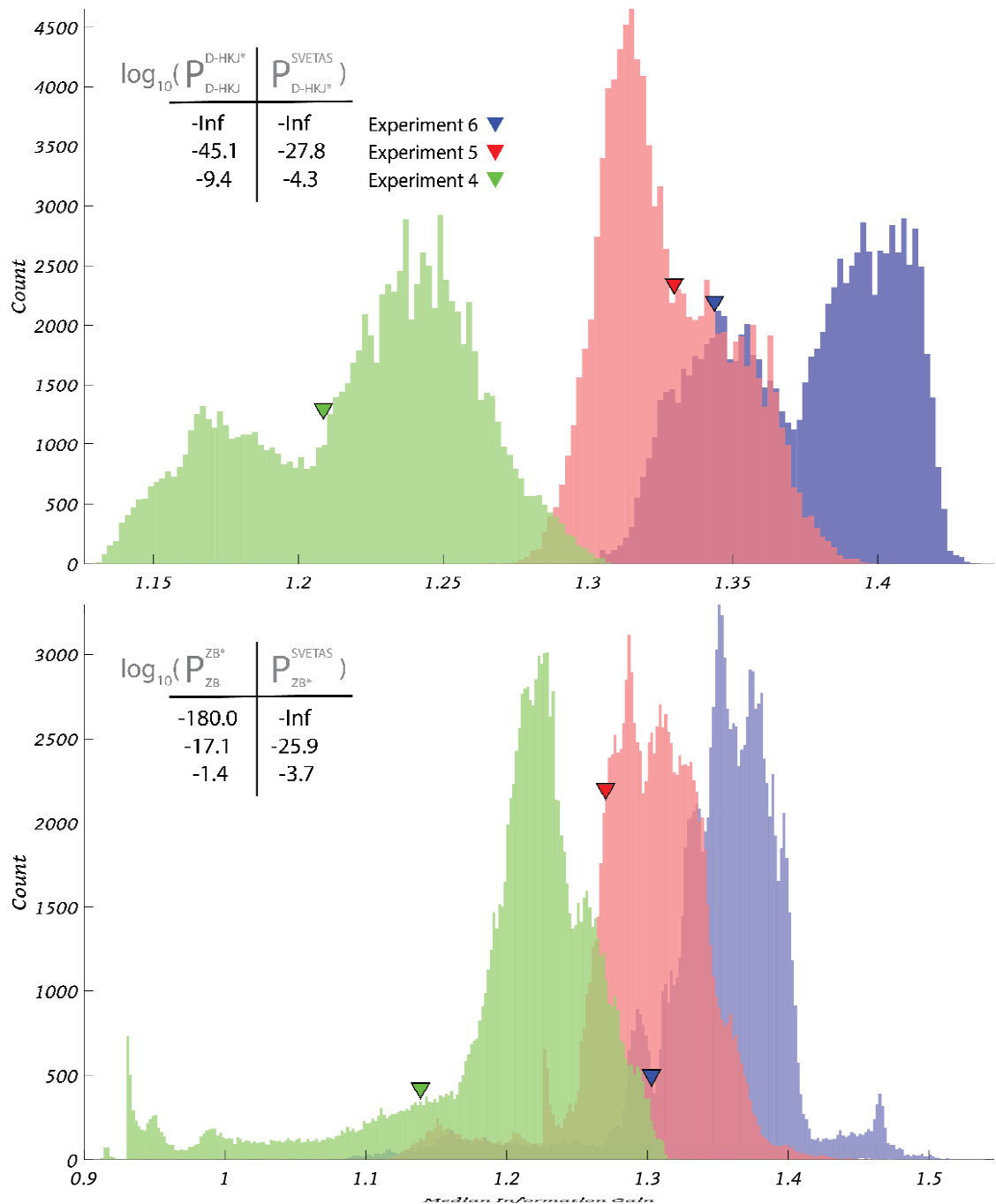

**Figure S2**: Distribution of the median information gain obtained for different values of declustering parameters of the Reasenberg's (top panel) and Zaliapin's (bottom panel) declustering algorithms, for the experiments 4, 5 and 6. Median is obtained from the sample information gain of each earthquake in all the 1 year long testing periods between 1999 and 2016; triangular markers indicate the median information gain obtained for the values of standard declustering parameters; $\log_{10}$ of p-values of the W-test comparing the sample information gain of the optimal models (D-HKJ* and D-ZB*) to the D-HKJ model and D-ZB model respectively are shown in the first column of inset tables; p-values of the W-test comparing the sample information gain of the SVETAS model to optimal models (D-HKJ* and D-ZB*) are shown in second column of the inset tables.